\documentclass[a4paper]{aa}

\usepackage{natbib}
\usepackage{graphicx}
\usepackage{txfonts}
\usepackage[bookmarks=false, colorlinks=true, citecolor=blue, linkcolor=blue]{hyperref}
\usepackage{upgreek}
\usepackage{pdflscape}
\usepackage{mathrsfs}

\newcommand{\JWST}{{JWST}}
\def\micron{\hbox{\,$\upmu$m}}
\newcommand{\Lsun}{\hbox{$L_{\rm \odot}$}}
\newcommand{\Msun}{\hbox{$M_{\rm \odot}$}}

\newcommand{\degree}{\ensuremath{^\circ}}
\newcommand\nodata{ ~$\cdots$~ }

\bibpunct{(}{)}{;}{a}{}{,}

\titlerunning{The nucleus of NGC~3256\,S. I. Anisotropic active galactic nucleus near-IR radiation field}
\authorrunning{Pereira-Santaella et al.}

\begin{document}

\defcitealias{Pereira2024CO}{PS24}

\title{The obscured nucleus of NGC~3256\,S} \subtitle{I. Anisotropic active galactic nucleus near-IR radiation field traced by CO $v$=1--0 with JWST\slash NIRSpec}

\author{Miguel~Pereira-Santaella\inst{\ref{inst1}}
\and Eduardo~Gonz\'alez-Alfonso\inst{\ref{inst2}}
\and Ismael~Garc\'ia-Bernete\inst{\ref{inst3}}
\and Andreas~Efstathiou\inst{\ref{inst9}}
\and Dimitra~Rigopoulou\inst{\ref{inst4},\ref{inst9}}
\and Giovanna~Speranza\inst{\ref{inst1}}
\and Niranjan~Thatte\inst{\ref{inst4}}
}

\institute{Instituto de F\'isica Fundamental, CSIC, Calle Serrano 123, 28006 Madrid, Spain \\
\email{miguel.pereira@csic.es}\label{inst1}
\and
Universidad de Alcal\'a, Departamento de F\'isica y Matem\'aticas, Campus Universitario, 28871 Alcal\'a de Henares, Madrid, Spain\label{inst2}
\and
Centro de Astrobiolog\'ia (CAB), CSIC-INTA, Camino Bajo del Castillo s/n, E-28692 Villanueva de la Ca\~nada, Madrid, Spain\label{inst3}
\and
Department of Physics, University of Oxford, Keble Road, Oxford OX1 3RH, UK\label{inst4}
\and
School of Sciences, European University Cyprus, Diogenes street, Engomi, 1516 Nicosia, Cyprus\label{inst9}
}

\abstract{
The southern nucleus of NGC~3256, a nearby luminous infrared (IR) interacting system ($d$=40\,Mpc; $L_{\rm IR}$=10$^{11.6}$\,\Lsun), hosts a deeply obscured low-luminosity active galactic nucleus (AGN) associated with a massive and fast bipolar collimated molecular outflow. \JWST\slash NIRSpec observations revealed bright spatially resolved ro-vibrational CO $v$=1--0 4.67\micron\ emission {at the base of} this molecular outflow {extending out to $\sim$300\,pc, which is} excited by the $\sim$4.7\micron\ IR radiation from the southern dusty core.
Using Monte Carlo radiative transfer models, we analyzed the CO $v$=1--0 band emission to determine the intensity of the $\sim$4.7\micron\ IR radiation field across the outflow {base}.
We find that the core spectral luminosity density at $\sim$4.7\micron\ as seen by the material in the outflow { base} is $\sim$2--4 times higher than that inferred along our line of sight, indicating an anisotropic near-IR radiation field.
Moreover, the slower than $d^{-2}$ decrease of the near-IR flux density across the outflow { base} suggests a collimated radiation field, for which we estimate a small half-opening angle (15--20\degree) and a diameter of the collimating structure of 15$\pm$5\,pc. This is slightly smaller than the diameter of the obscuring envelope (25$\pm$5\,pc).
Combined with the expansion of this obscuring envelope, traced by the blueshifted CO $v$=1--0 absorption toward the nucleus, this suggests that the AGN is beginning to clear its environment.
We estimate $N_{\rm H}$$\sim$10$^{22-23}$\,cm$^{-2}$ for NGC~3256\,S, which is similar to that of Type 2 AGNs. However, in contrast to Type 2 AGNs, a high covering factor of the obscuring material, possibly sustained by inflows linked to the galaxy merger and similar to that expected in deeply obscured compact obscured nuclei (CONs), is needed to explain the absence of extended high-ionization emission. Therefore, NGC~3256\,S shares properties with CONs and Type 2 AGNs.
}


\maketitle

\section{Introduction}\label{sec:intro}

\begin{figure*}
\centering
\includegraphics[width=\textwidth]{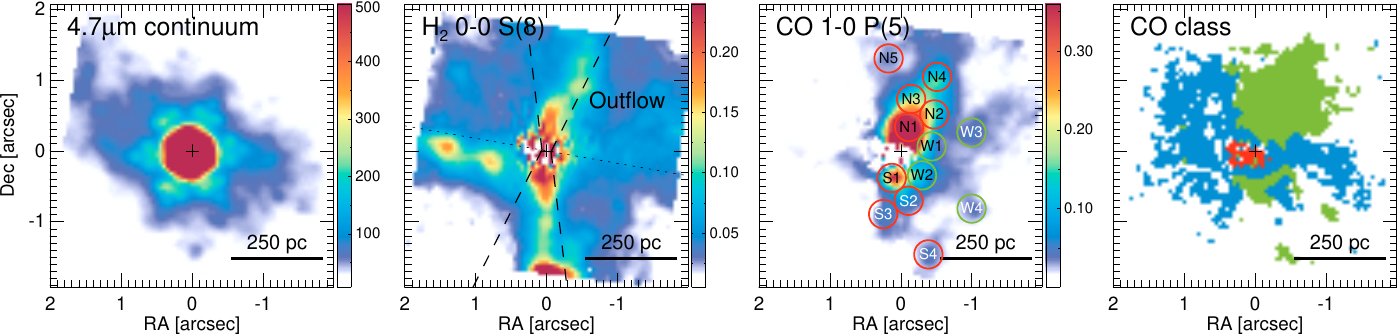}
\caption{JWST\slash NIRSpec maps of NGC~3256\,S. First panel: 4.7\micron\ continuum map. The units are MJy\,sr$^{-1}$. Second and third panels: Maps of the H$_2$ $v$=0--0 S(8) 5.05\micron\ and CO $v$=1--0 P(5) 4.71\micron\ lines, respectively. The units are 10$^{-19}$\,erg\,cm$^{-2}$\,s$^{-1}$\,arcsec$^{-2}$. 
The dashed lines in the second panel delineate the molecular outflow area. The dotted line marks the direction perpendicular to the outflow axis (approximate galaxy disk location). The position and size of the apertures described in Sect.~\ref{s:data_reduction} are indicated the third panel.
Fourth panel: Classification of the CO $v$=1--0 band according to the absorption or emission of the P and R branches. The red area marks regions in which both branches are in absorption. The blue area indicates regions in which the R branch is in absorption and the P branch in emission based on the R(2) and P(5) lines (see Sect.~3.2 of \citetalias{Pereira2024CO}). The green areas are regions in which both branches are in emission. The black cross in all the panels marks the position of the 4.7\micron\ near-IR continuum peak at R.A. (ICRS) $=$ 10h27m51.22s Decl. (ICRS) $=$ $-$43d54m19.2s.
\label{fig_line_maps}}
\end{figure*}

{
Mergers of gas-rich galaxies funnel large quantities of gas and dust into their central regions, triggering intense star formation and active galactic nuclei (AGNs; e.g., \citealt{Sanders96, Hopkins2013, RodriguezMontero2019}).
This activity initially occurs in heavily dust-obscured environments, where ultraviolet and optical emission from young stars and the AGN is absorbed by dust and is re-emitted in the infrared (IR), leading to IR-bright galaxies with $L_{\rm IR}>$10$^{11-12}$\,\Lsun, i.e., luminous and ultraluminous IR galaxies (LIRGs and ULIRGs).
Among the most deeply embedded systems are compact obscured nuclei (CONs), with $N_{\rm H}$>10$^{24-25}$\,cm$^{-2}$. CONs are characterized by their exceptionally compact and strong mid-IR radiation densities, which are capable of exciting submillimeter HCN-vib emission \citep{Falstad2021, GarciaBernete2025, GonzalezAlfonso2026}.
As star formation- and AGN-driven feedback clears the obscuring dust and gas, ULIRGs, the most luminous systems, are thought to transition into quasars (e.g., \citealt{Snyder2013, Kawaguchi2020}). This clearing is consistent with the ubiquitous detection of massive molecular outflows in ULIRGs \citep{Sturm2011, Veilleux2013, Cicone2014, GonzalezAlfonso2017, Pereira2018, Lutz2020, Lamperti2022}.

We here focus on the nearby LIRG NGC~3256 ($d$=40\,Mpc; $L_{\rm IR}$=4$\times$10$^{11}$\,$L_\odot$), which hosts an obscured AGN in its southern nucleus as well as fast massive molecular outflows. Thus, owing to its proximity, it provides a unique opportunity to investigate the transition from a dust-obscured into an unobscured AGN phase in a gas-rich merger in detail.

 The NGC~3256 system is an advanced merger (projected nuclear separation of $<$1\,kpc).
The northern component is dominated by a strong starburst with a star formation rate of $\sim$15\,$M_\odot$\,yr$^{-1}$ \citep{Lira2008}, which is not investigated in this paper.
In contrast, the southern component, NGC~3256\,S, hosts an extremely dust-embedded nucleus (e.g., \citealt{Pereira2010, DiazSantos2010}) powered by a low-luminosity obscured AGN ($L_{\rm 2-10\,keV}\sim10^{40}$\,erg\,s$^{-1}$), as suggested by multiwavelength modeling of near-IR, mid-IR, and X-ray observations \citep{Ohyama2015}.
Similar to CONs, no high-ionization transitions or direct X-ray emission are detected in NGC 3256 S from the AGN (e.g., \citealt{Lira2002, AAH2012a, Pereira2011, Pereira2024Mg4, GarciaBernete2025}).
NGC~3256\,S also hosts a large-scale ($\sim$1\,kpc) high-velocity ($\sim$1000\,km\,s$^{-1}$) collimated bipolar molecular outflow, in which hot ($T$$\sim$2000\,K), warm ($T$$\sim$300\,K), and cold ($T$=15--50\,K) molecular gas have been detected using H$_2$ and CO lines \citep{Sakamoto2014, Emonts2014, Bohn2024, Pereira2024CO}, with the cold molecular phase accounting for the great majority (>90\%) of the outflow mass. Similar collimated molecular outflows have been detected in other (U)LIRGs classified as CONs (e.g., ESO~320-G030, \citealt{Pereira2016b, Gorski2024}; Arp~220, \citealt{BarcosMunoz2018}; and Zw\,049-057, \citealt{Lankhaar2024}).
Therefore, NGC~3256\,S exhibits some properties associated with CONs, although its nuclear $N_{\rm H}$ and mid-IR radiation density are likely lower than that of CONs, so no bright HCN-vib emission is produced.

In \citet{Pereira2024CO}, hereafter \citetalias{Pereira2024CO}, we analyzed the CO \hbox{$v$=1--0} 4.67\micron\ ro-vibrational band emission observed by the Near-Infrared Spectrograph (NIRSpec) aboard the James Webb Space Telescope (JWST). {This ro-vibrational emission is  spatially coincident with the {base of the} molecular outflow detected in the Atacama Large Millimeter Array (ALMA) data and shows consistent kinematics, suggesting that this emission is associated with the outflowing molecular gas of NGC~3256\,S (see \citetalias{Pereira2024CO}).}
This CO emission is produced by resonant scattering of the AGN near-IR light, analogous to the mechanism observed in Orion \citep{GonzalezAlfonso1998, GonzalezAlfonso2002}.
Using the large velocity gradient approximation, we modeled the line flux ratios of the P($J$+2) and the R($J$) transitions of the band, which share the same upper level. From these models, we constrained the physical conditions of the gas in the outflow { base} and obtained these results: (a) the CO \hbox{$v$=1--0} emission is excited by the intense near-IR illumination from the AGN up to $\sim$300\,pc; (b) there is a significant temperature gradient in the molecular gas from $>$40\,K at the nucleus to $<$15\,K at 300\,pc; and (c) the gas in the outflow { base} has a low $\alpha_{\rm CO}$ conversion factor of $\sim$(0.40--0.61)$\times\frac{3.2 \times10^{-4}}{[{\rm CO}\slash {\rm H_2}]}$\,$M_\odot$ (K km\,s$^{-1}$\,pc$^2$)$^{-1}$.

This is the first paper in a two-paper series. We extend our analysis of the molecular gas in the outflow { base} applying a Monte Carlo radiative transfer (MC RT) model (see, e.g., \citealt{Steinacker2013}). This approach allows us to model not only line ratios, but also the absolute luminosities of the band. In this first paper, we investigate the CO \hbox{$v=$1--0} emission, and in the second part (Pereira-Santaella et al. in prep.), we discuss the H$_3^+$ $v_2$=1--0 band detected in emission for the first time in the interstellar medium in this outflow \citep{Pereira2024H3p}.

This paper is organized as follows: 
The data reduction and aperture spectral extraction are presented in Sect.~\ref{s:data_reduction}. 
Sect.~\ref{s:results} describes the MC RT models and the fitting of the observed spectra.
In Sect.~\ref{s:discussion} we discuss the inferred properties of the radiation field and nuclear obscuring material.
Finally, in Sect.~\ref{s:conclusions}, we summarize the main results of the paper.

}

\section{Data reduction}\label{s:data_reduction}

We used high-spectral-resolution ($R$$\sim$1900--3600) \JWST\slash NIRSpec integral-field spectroscopy obtained with the grating\slash filter pair G395H\slash F290LP (2.87--5.27\micron).
These observations are part of the director's discretionary early release science (DD-ERS) program 1328 (PI: L. Armus and A. Evans). The field of view of these data is 3\farcs1$\times$3\farcs1 and the angular resolution is $\sim$0\farcs2. 
The same observations were analyzed in \citetalias{Pereira2024CO}, and we therefore followed the same reduction steps, but used an updated version of the pipeline (1.12.4) and calibration context (1254) and a smaller spaxel size of 0\farcs05 for the final cubes to better recover the spatial information.
More details on the data reduction can be found in \citetalias{Pereira2024CO} and \citet{GarciaBernete2024_ice}.

In addition, we combined the \JWST\slash NIRSpec data with CO $J$=1--0 (115.27\,GHz) ALMA observations of NGC~3256 from program 2018.1.00223.S (PI: K. Sakamoto). This dataset was also used in \citetalias{Pereira2024CO}, and its standard data reduction is described in more detail there. The beam full width at half maximum (FWHM) of this ALMA cube is 0\farcs39$\times$0\farcs37.

As discussed in \citetalias{Pereira2024CO}, the CO $v$=1--0 band in NGC 3256 S is seen in emission along the outflow { base} in absorption toward the nucleus, while the P branch is in emission and the R branch in absorption (P-R asymmetry) along the disk of the galaxy (see the fourth panel of Fig.~\ref{fig_line_maps}).
The MC RT models we present are applicable to areas in which the CO band appears in pure emission, and we thus selected multiple regions mostly along the outflow axis (see the third panel of Fig.~\ref{fig_line_maps}).

We used apertures with a diameter of 0\farcs4 ($\sim$80\,pc) and subtracted the contribution produced by the wings of the bright nuclear point-spread function following the method described in sect.~3.3 of \citetalias{Pereira2024CO}. We also used these apertures to extract the rotational CO $J$=1--0 emission from the ALMA cubes. {We note that the apertures we used are smaller and encompass slightly different regions than those used in \citetalias{Pereira2024CO}. We provide an approximate correspondence between the apertures in Table~\ref{tbl_bestfit}.}
The ALMA and JWST data cubes were aligned based on the positions of the 5\micron\ near-IR and the 115\,GHz millimeter continua.

\section{Analysis and results}\label{s:results}

The stellar disk of NGC~3256\,S is seen almost edge-on from our line of sight, while the molecular outflow lies approximately in the plane of the sky (e.g., \citealt{Sakamoto2014}).
This orientation is ideal for the study of the CO $v$=1--0 emission { at base of the outflow}, since it limits the contribution of the disk to the spectra of the outflow (see Fig.~\ref{fig_line_maps}). The CO $v$=1--0 band is excited by the bright near-IR continuum emission produced by the warm dust ($T_{\rm d}\sim600$\,K; \citealt{Ohyama2015}) in the dust-embedded nucleus of NGC~3256\,S.
This near-IR continuum illuminates the molecular gas in the outflow { base}, exciting the $v$=1 vibrational levels of the CO molecules, which subsequently decay through the observed $v$=1--0 emission (see Fig.~6 of \citetalias{Pereira2024CO}).

In \citetalias{Pereira2024CO}, we showed that the low-$J$ lines of the band (up to $J_{\rm lo}\sim$3--4)  are optically thick. To model the observed intensities of all the lines of band, a detailed radiative transfer treatment is therefore required. 
To achieve this, we developed an MC RT model (Sect.~\ref{ss:mcrt}) that is sensitive to the different optical depths of the lines in the band and can predict the expected luminosity of each line.
We used the MC RT model predictions to fit the observed band emission and constrain the physical conditions  of the outflowing gas and the near-IR radiation field (Sect.~\ref{ss:fit}).

\subsection{MC RT model for the CO $v$=1--0 band}\label{ss:mcrt}

\begin{figure*}
\centering
\includegraphics[width=0.9\textwidth]{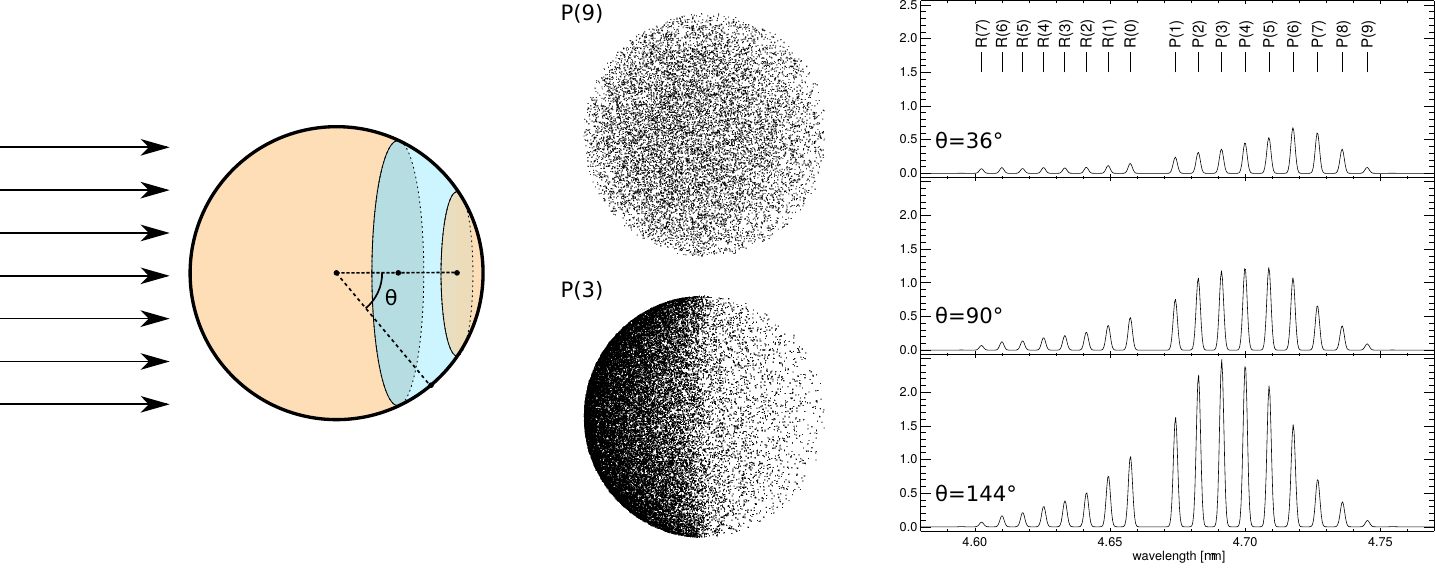}
\caption{Left panel: Assumed geometry and side illumination (black arrows) of the spherical clouds by the near-IR radiation field. The observer viewing angle, $\theta$, and the corresponding spherical segment for this axisymmetric geometry are shown in blue. Middle and right panels: MC RT results for a model with $T_{\rm kin}$=20\,K, $n_{\rm H_2}$=10$^4$\,cm$^{-3}$, and $\langle N_{\rm CO}\rangle\slash \Delta {\rm v}$=10$^{18.3}$\,cm$^{-2}$\slash (km\,s$^{-1}$). The middle panel shows the last scattering location of the rays before they escape the cloud for an optically thick line (bottom; P(3)) and an optically thinner line (top; P(9)). For the thinner line, the distribution is almost homogeneous and isotropic, while for the thicker line, the emission is mostly produced at the illuminated side of the cloud, thus highly anisotropic. The right panels show the simulated CO $v$=1--0 band for three viewing angles in arbitrary flux units.
\label{fig_mcrt}}
\end{figure*}

We modeled the scattering of the nuclear near-IR radiation by CO molecules in the outflow using an MC RT approach. This is appropriate for modeling the anisotropic emission expected from resonant scattering of the incident near-IR radiation because the opacity of the low-$J$ lines of the CO $v$=1--0 band is high \citepalias{Pereira2024CO}.

To simulate the illumination of the molecular gas in the outflow by the compact IR core, we assumed homogeneous spherical clouds with half-hemisphere irradiated by photons following parallel tracks (see Fig.~\ref{fig_mcrt} left). This is a highly symmetric geometry, so a simple MC RT calculation is sufficient to achieve accurate results (see, e.g., \citealt{Steinacker2013}). We assumed that the cloud had a Gaussian velocity distribution with a dispersion, $\sigma_{\rm v}\sim 0.425 \Delta {\rm v}$, where $\Delta {\rm v}$ represents the FWHM, and a CO column density along its diameter, $N_{\rm CO}^{\rm max}$, that corresponds to an average CO column density of $\langle N_{\rm CO}\rangle=2\slash 3N_{\rm CO}^{\rm max}$ for the cloud.

The steps of the calculation are the following: (1) we defined the initial wavelength and spatial coordinates of the photons. The wavelength grid covers a small range around each ro-vibrational transition (see below) to account for  the varying optical depth across the velocity resolved line profiles. The photons were uniformly distributed in a circle with a radius matching that of the sphere, and located just outside its surface. They were also distributed in the wavelength grid following the intensity of a blackbody with $T_{\rm d}$ to simulate the spectral energy distribution (SED) of the near-IR radiation field. $T_{\rm d}$ is not well constrained by the data ($\sim$300--600\,K are compatible with the near-IR continuum and CO band), so we assumed a fixed intermediate $T_{\rm d}$=400\,K, which is also consistent with the observed nuclear continuum in our line of sight.
(2) The optical depth, $\tau$, traveled by the photon before it interacts with a CO molecule was then randomly generated using an exponential distribution (eq.~25 of \citealt{Steinacker2013}). This $\tau$ was then compared with the optical depth from the location of the photon to the surface of the cloud, $\tau_{\rm s}$.
(3) If $\tau>\tau_{\rm s}$, the photon escapes the system and its wavelength and escape direction were recorded. Otherwise, the absorption and re-emission of the photon by a CO molecule was simulated. According to its energy, the absorbed photon excites a CO molecule to a specific $v$=1 level. Then, a new wavelength is assigned to the re-emitted photon based on the transition probabilities of the possible decay paths (P and R branches of the $v$=1--0 band, and pure rotational transitions within $v$=1).
Finally, assuming that the CO molecules emit isotropically, the re-emitted photon was assigned a new propagation direction sampled uniformly in solid angle, corresponding to complete redistribution in angles. Steps 2 and 3 were repeated until all photons escape the cloud.

For step 2, we needed to compute $\tau_{\rm s}$ as function of $N_{\rm CO}^{\rm max}$. The optical depth depends on the population of the energy levels of the CO molecules (see, e.g., Eqs.~6--9 of \citetalias{Pereira2024CO}). Based on the analysis in \citetalias{Pereira2024CO}, the great majority of CO { at the base of the} outflow is in the ground vibrational state, and its excitation is driven by collisions with H$_2$ and not by IR pumping.
Therefore, we assumed that the CO level populations only depend on the kinetic temperature, $T_{\rm kin}$, the H$_2$ density, $n_{\rm H_2}$, and $N_{\rm CO}^{\rm max}\slash \Delta{\rm v}$, which were frozen during the MC RT calculations.
Under these assumptions, we created a grid of models using the non local thermodynamic equilibrium (non-LTE) code \textsc{RADEX} \citep{vanderTak2007} with $T_{\rm kin}$=12--125\,K, $n_{\rm H_2}$=10$^2$--10$^6$\,cm$^{-3}$, and $N_{\rm CO}^{\rm max}\slash \Delta{\rm v}$=10$^{17}$--10$^{20}$\,cm$^{-2}$\slash (km\,s$^{-1}$) with 25 log steps for $T_{\rm kin}$ and $N_{\rm CO}^{\rm max}\slash \Delta{\rm v}$ and 10 log steps for $n_{\rm H_2}$, due of the reduced sensitivity of the models to $n_{\rm H_2}$ variations. In total, the grid had 6250 models. For the \textsc{RADEX} calculation of the $v$=0 level populations, we used the spectroscopic parameters of CO from \citet{Goorvitch1994} and the collisional coefficients from \citet{Yang2010}.

Using the \textsc{RADEX} grid of models, we estimated the optical depth at the line center, $\tau_{\rm s}(0)$, for each vibrational CO transition. Some of these CO transitions have high opacities, so the core of their profiles are saturated while their wings are optically thin.
To account for this, we defined a grid of velocities between 0 and $A\sigma_{\rm v}$, with $A$ between 5 and 7 depending on $\tau_{\rm s}(0)$, to cover the line wings down to $\tau_{\rm s}(v)<$0.05, and a sampling of $\sigma_{\rm v}\slash$2. Since the line profiles are symmetric, we only computed half of the profile with the MC RT. So for each transition, we simulated between 10 and 14 velocity channels.
Between 80000 and 110000 model photons were assigned to each velocity channel of each transition according to the blackbody SED. A total of $\sim$10 to 40 million of model photons were traced for each MC RT simulation, depending on $N_{\rm CO}^{\rm max}$.

The optically thin lines are re-emitted isotropically by the cloud, so their intensities do not depend on the observer line of sight. However, the intensity of the optically thick lines is highly dependent on the viewing angle ($\theta$), with stronger emission facing the illuminated side of the cloud ($\theta$>90\,deg) and fainter emission at the opposite side ($\theta$<90\,deg; middle panels of Fig.~\ref{fig_mcrt}).
Therefore, using the recorded escape direction of the rays, we computed the angular dependence of the cloud emission. We defined 20 equal angular bins for observing angles $\theta$ between 0 and 180\,deg (see Fig.~\ref{fig_mcrt}, left). The solid angle of each bin is different, so we normalized the number of rays in each one by the solid angle of the corresponding spherical segment. 
Then, for each transition, we calculated the fraction of the total emission that falls within each angular bin. Using this fraction, we derived the integrated luminosity of any line $i$ produced by the cloud, $L_{{\rm line},i}$, from the observed flux in any direction $\theta$.

We also used these models to derive the monochromatic luminosity of the near-IR source around 4.7\micron, $L_{\nu}$, that illuminates the clouds based on the scattered CO $v$=1--0 band. The total continuum power density incident on a cloud, $P_\nu$, is 
\begin{equation}
P_\nu = A_{\rm cloud}\times F_\nu(d),
\label{eq_Pnu}
\end{equation}
where $A_{\rm cloud}$ is the cross-section area of the spherical cloud, and $F_\nu(d)$ is the flux density from the central IR source at a distance $d$.
For isotropic emission, $F_\nu(d) = L_\nu \slash ( 4\pi d^2 )$.
For these models, the radius of the cloud, $r_{\rm cloud}$, can be expressed as a function of the \textsc{RADEX} parameters,
\begin{equation}
r_{\rm cloud} = \frac{1}{2}\left(\frac{N_{\rm CO}^{\rm max}}{\Delta {\rm v}}\right) \frac{\Delta {\rm v}}{n_{\rm CO}},
\end{equation}
where $n_{\rm CO}$ is the CO number density. The number of CO molecules in the cloud, $\mathcal{N}$, is $4\slash 3\pi r_{\rm cloud}^3 n_{\rm CO}$, so $A_{\rm cloud}$ can be written as
\begin{equation}
A_{\rm cloud}=\pi r_{\rm cloud}^2 = \frac{3 \mathcal{N}}{2 \Delta {\rm v}} \left(\frac{N_{\rm CO}^{\rm max}}{\Delta {\rm v}}\right)^{-1}.
\label{eq_A}
\end{equation}
From the MC RT models, we obtained the ratio, $r_i$, of the continuum power density integrated over a velocity bin, $\Delta {\rm v} P_\nu$, and the line luminosity, $r_i=\Delta {\rm v} P_\nu \slash L_{{\rm line},i}$. When this result is combined with Eqs. \ref{eq_Pnu} and \ref{eq_A}, the near-IR flux density exciting the CO $v$=\hbox{1--0} band is
\begin{equation}
F_\nu(d) = \left(\frac{N_{\rm CO}^{\rm max}}{\Delta {\rm v}}\right)\frac{1}{\mathcal{N}}\frac{2}{3} r_{\rm i} L_{\rm line, i}\,,\label{eq_lagn}
\end{equation}

\noindent which can be determined from the observed luminosity of each line of the CO band. The number of CO molecules, $\mathcal{N}$, was self-consistently obtained from the ALMA CO(1--0) emission using the emission predicted for this line by each \textsc{RADEX} model (see also \citetalias{Pereira2024CO}).

\begin{figure*}
\centering
\includegraphics[width=0.97\textwidth]{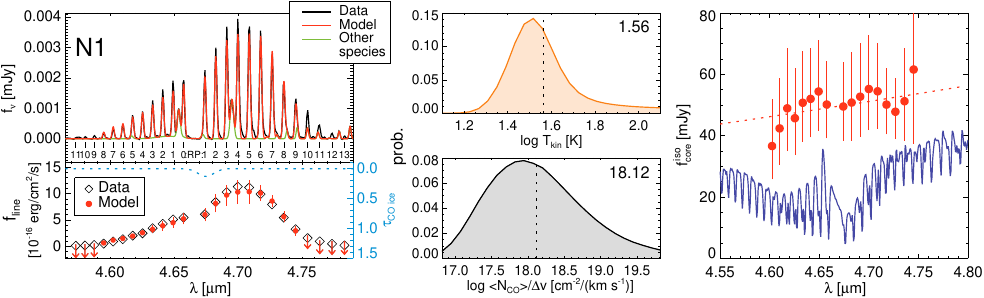}
\includegraphics[width=0.97\textwidth]{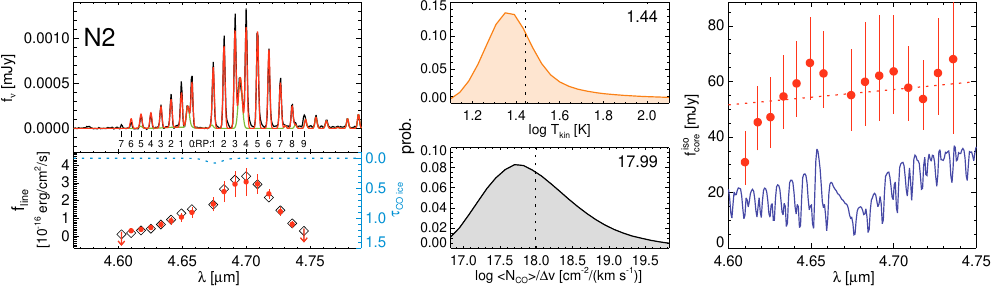}
\includegraphics[width=0.97\textwidth]{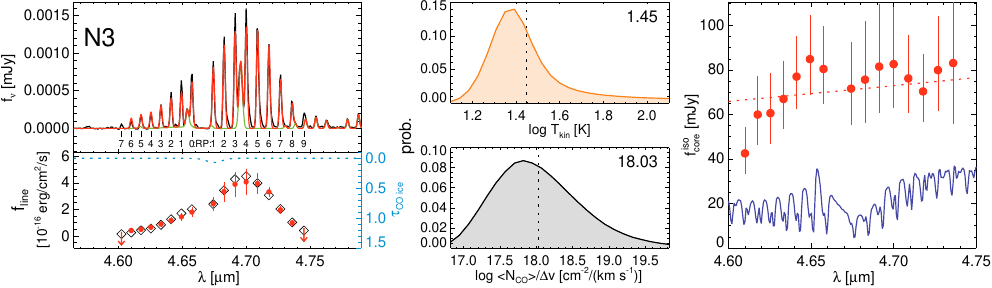}
\includegraphics[width=0.97\textwidth]{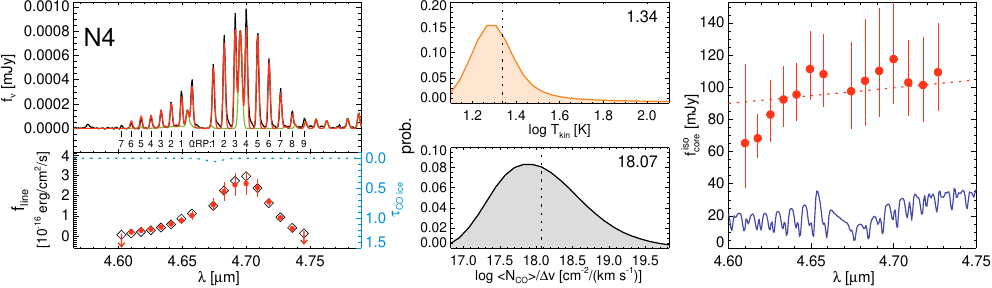}
\caption{MC RT fitting results for each region. The name of the region is indicated in the top-left panel. Top left: Continuum-subtracted \JWST\slash NIRSpec spectrum (solid black line). The red line is the  best-fit CO $v$=1--0 model plus the contribution from the other species, indicated as a solid green line (see Sect.~\ref{ss:fit}). Bottom left; Observed fluxes of the CO $v$=1--0 lines (black diamonds) as a function of the wavelength. The flux of the best-fit model is represented by the red circles and downward arrows for the upper limits. The dashed blue line shows the fitted optical depth for CO ice band at 4.67\micron\ (right axis). Middle panels: Probability density distributions for $\log T_{\rm kin}$ (top panel in orange) and $\log \langle N_{\rm CO}\rangle \slash \Delta{\rm v}$ (bottom panel in black). The median value is indicated in the top right corner of the panel and shown with a vertical dotted line. Right panel: Hypothetical core flux derived from the MC RT model for each CO $v$=1--0 line assuming that the core emission is isotropic (red circles). The dashed red line represents the scaled emission of blackbody at 400\,K. For reference, the nuclear spectrum of NGC~3256\,S measured with \JWST\slash NIRSpec is shown in blue. The discrepancy between the observed nuclear flux (blue line) and that derived from the MC RT model (red points) suggests that the core does not emit isotropically (see Sect.~\ref{ss:anisotropy}).
\label{fig_bestfit}}
\end{figure*}

\begin{table*}
\caption{Properties of the regions.}
\label{tbl_bestfit}
\centering
\begin{small}
\begin{tabular}{lcccccccccc}
\hline \hline
\\
Region & $d$\,$^{(a)}$ & log\,$L^\prime_{\rm CO(J=1-0)}$\,$^{(b)}$ & log\,$L_{\rm band}$\,$^{(c)}$ & log\,$F_\nu$\,$^{(d)}$ & log$T_{\rm kin}$\,$^{(e)}$ & log,$\langle N_{\rm CO}\rangle\slash \Delta {\rm v}$\,$^{(e)}$ & $\tau_{\rm CO\,ice}$\,$^{(f)}$ & \citetalias{Pereira2024CO} \\
& (pc) & (K\,km\,s$^{-1}$\,pc$^2$) & (erg\,s$^{-1}$) &  (Jy) & (K) & (cm$^{-2}$\slash (km\,s$^{-1}$)) & & region\,$^{(g)}$ \\
\hline
N1 & 68 & 7.08 & 39.29 & 10.21 $\pm$ 0.03 & 1.56 $\pm$ 0.16 & 18.12 $\pm$ 0.63 & 0.13 &   N1  \\
N2 & 136 & 6.95 & 38.65 & 9.66 $\pm$ 0.03 & 1.44 $\pm$ 0.20 & 17.99 $\pm$ 0.63 & 0.08 &   N1  \\
N3 & 140 & 7.02 & 38.78 & 9.74 $\pm$ 0.03 & 1.45 $\pm$ 0.18 & 18.03 $\pm$ 0.59 & 0.07 &   N1  \\
N4 & 221 & 6.96 & 38.54 & 9.48 $\pm$ 0.03 & 1.34 $\pm$ 0.16 & 18.07 $\pm$ 0.59 & 0.05 &   N2  \\
N5 & 244 & 6.71 & 38.05 & 9.20 $\pm$ 0.03 & 1.23 $\pm$ 0.16 & 17.84 $\pm$ 0.55 & 0.13 &   N3  \\
S1 & 75 & 6.96 & 39.09 & 10.27 $\pm$ 0.03 & 1.59 $\pm$ 0.16 & 18.26 $\pm$ 0.67 & 0.34 &   \nodata  \\
S2 & 137 & 6.63 & 38.37 & 9.77 $\pm$ 0.03 & 1.43 $\pm$ 0.16 & 18.37 $\pm$ 0.71 & 0.16 &   \nodata  \\
S3 & 174 & 6.55 & 38.16 & 9.58 $\pm$ 0.03 & 1.38 $\pm$ 0.20 & 18.32 $\pm$ 0.79 & 0.14 &   S1  \\
S4 & 288 & 6.65 & 38.04 & 9.32 $\pm$ 0.03 & 1.29 $\pm$ 0.16 & 18.36 $\pm$ 0.73 & 0.23 &   S2  \\
W1 & 91 & 6.96 & 38.50 & 9.81 $\pm$ 0.03 & 1.41 $\pm$ 0.15 & 18.70 $\pm$ 0.72 & 0.34  &   \nodata  \\
W2 & 93 & 6.93 & 38.59 & 9.70 $\pm$ 0.02 & 1.44 $\pm$ 0.13 & 18.53 $\pm$ 0.67 & 0.48  &   \nodata  \\
W3 & 202 & 6.85 & 38.09 & 9.11 $\pm$ 0.03 & 1.27 $\pm$ 0.15 & 18.67 $\pm$ 0.72 & 0.10 &   N4  \\
W4 & 250 & 6.70 & 37.99 & 9.19 $\pm$ 0.03 & 1.34 $\pm$ 0.21 & 18.04 $\pm$ 0.75 & 0.14 &   S3  \\
\hline
\end{tabular}
\end{small}
\tablefoot{
$^{(a)}$ Projected separation between the nucleus and the region. No correction for inclination was applied because the outflow is close to the plane of the sky.
$^{(b)}$ Luminosity of the CO(1--0) 115.27\,GHz rotational line calculated using eq.~3 of \citet{Solomon2005}.
$^{(c)}$ Integrated luminosity of all the lines of the ro-vibrational CO $v$=1--0 4.67\micron\ band.
$^{(d)}$ Spectral continuum flux density at $\sim$4.7$\micron$ exciting the ro-vibrational CO emission derived from the MC RT modeling (Sect.~\ref{ss:fit}).
$^{(e)}$ Mean $T_{\rm kin}$ and $\langle N_{\rm CO}\rangle\slash \Delta {\rm v}$ derived from the MC RT modeling (Sect.~\ref{ss:fit}).
$^{(f)}$ Measured optical depth of the CO ice absorption at $\sim$4.67\micron\ affecting the CO band emission (Sect.~\ref{ss:fit}).
{$^{(g)}$ Approximate correspondence with the larger apertures (0\farcs5 diameter) defined in \citetalias{Pereira2024CO}.}
}
\end{table*}

\subsection{Best-fit parameters}\label{ss:fit}

To estimate the best-fit parameters for the regions { at the base of the outflow}, we first measured the observed flux of all the CO $v$=1--0 lines. Following \citetalias{Pereira2024CO}, we subtracted a baseline continuum and included it in the line-fitting procedure lines from other species (the $^{13}$CO $v$=1--0 band, the 4.65\micron\ deuterated polycyclic aromatic hydrocarbon (D-PAH) band, H$_2$ and \ion{H}{I} recombination lines, and [\ion{K}{iii}]4.62\micron) that lie in this wavelength range (4.55--4.80\micron). This contribution from other species is shown in the first column panel of Fig.~\ref{fig_bestfit} for each region as a solid green line.

Then, we calculated the $\chi^2$ for each of the 6250 MC RT models for viewing angles within the range stated below by comparing the predicted fluxes of the CO $v$=1--0 lines with the observed ones. 
For each model, we also computed the expected $F_\nu(d)$ illuminating the molecular cloud for each CO $v$=1--0 line flux (Eq.~\ref{eq_lagn}). \citet{Sakamoto2014} estimated that the outflow lies close to the plane of the sky with an inclination $i$ of $\sim$80$^{+5}_{-10}$\degree. This corresponds to viewing angles $\theta$, as defined in Sect.~\ref{ss:mcrt} and Fig.~\ref{fig_mcrt}, of the outflowing clouds in the blue- and redshifted sides of the outflow from 70\degree\ to 110\degree, and we thus only considered lines of sight within this range. Finally, we assigned a probability proportional to $\exp(-\chi^2)$ to each model and determined the marginalized probability density functions for the model parameters and other derived physical quantities such as the predicted CO $v$=1--0 fluxes and $F_\nu(d)$.
Predicted line fluxes with a significance below 3$\sigma$ were treated as upper limits.

During this process, we noted that the best-fit models systematically underestimated the fluxes of the P(1) and P(2) transitions at 4.67--4.68\micron, in particular in region W2. This is likely caused by the CO ice absorption band at $\sim$4.67\micron\ \citep{Boogert2015} that reduces the fluxes of the lines.
This is plausible since the CO ice band is indeed detected in the nuclear spectrum of NGC~3256\,S (\citetalias{Pereira2024CO}) and extended ice absorptions from other species such as H$_2$O and CO$_2$ are widespread in local (U)LIRGs, including NGC~3256\,S \citep{Speranza26}. Therefore, we modeled the optical depth of this absorption with a Gaussian profile centered at 4.675\micron\ with $\sigma$=0.006\micron, whose peak optical depth was optimized iteratively to minimize the systematic discrepancy between the predicted and observed P(1) and P(2) fluxes.

The results from this fit are shown in Figs.~\ref{fig_bestfit} and \ref{fig_bestfit_apx} and are summarized in Table~\ref{tbl_bestfit}.
The predicted and observed CO band fluxes for all regions agree very well (left panels of Figs.~\ref{fig_bestfit} and \ref{fig_bestfit_apx}).
The probability distributions obtained for $T_{\rm kin}$ and $\langle N_{\rm CO}\rangle \slash \Delta{\rm v}$ are also shown in the middle panels of Fig.~\ref{fig_bestfit} and \ref{fig_bestfit_apx}. $T_{\rm kin}$ follows the same trend as reported in \citetalias{Pereira2024CO} with decreasing temperatures, from 40\,K to 20\,K, for increasing nuclear distances between 70 and 300\,pc {(see Fig.~12 of \citetalias{Pereira2024CO})} . No clear trend is observed for $\langle N_{\rm CO}\rangle \slash \Delta{\rm v}$, which varies between 10$^{18.0}$ and 10$^{18.5}$\,cm$^{-2}$\slash (km\,s$^{-1}$), although the values in the southern regions are higher by $\sim$0.3\,dex. The obtained $\langle N_{\rm CO}\rangle \slash \Delta{\rm v}$ are consistent within the uncertainties with those estimated using the two-temperature model in \citetalias{Pereira2024CO}. $n_{\rm H_2}$ it not shown since it is not well constrained by these data and its distribution is almost flat.

The right panels of Fig.~\ref{fig_bestfit} compare the actual nuclear spectrum measured with \JWST\slash NIRSpec (in blue) with the hypothetical flux expected at Earth derived using the MC RT model for each CO $v$=1--0 line (in red), assuming that the core emission is isotropic (i.e., $f^{\rm iso}_{\rm core} = 4\pi d^2 F_\nu(d)/(4 \pi D^2)$ with $D$=40\,Mpc and $d$ the projected separation between the nucleus and the center of each aperture). The molecular outflow in NGC~3256\,S is very close to the plane of the sky ($i$$\sim$80\degree; \citealt{Sakamoto2014}), so using the inclination corrected $d$ would increase $f^{\rm iso}_{\rm core}$ only by $\sim$10\%.
Except for W1, W2, and W3, which lie relatively far from the outflow axis, the core fluxes derived from the modeling exceed the observed nuclear emission by a factor of 2--4 across all regions.
This discrepancy suggests that the near-IR continuum emission of the nucleus is anisotropic, with lines of sight along the molecular outflow exposed to an enhanced near-IR radiation field.
{As expected, the modeled core flux increases with wavelength, consistent within the uncertainties with the assumed $T_{\rm d}$ of 400\,K.}

\section{Discussion}\label{s:discussion}

\subsection{Anisotropic nuclear near-IR emission}\label{ss:anisotropy}

\begin{figure}
\centering
\includegraphics[width=0.4\textwidth]{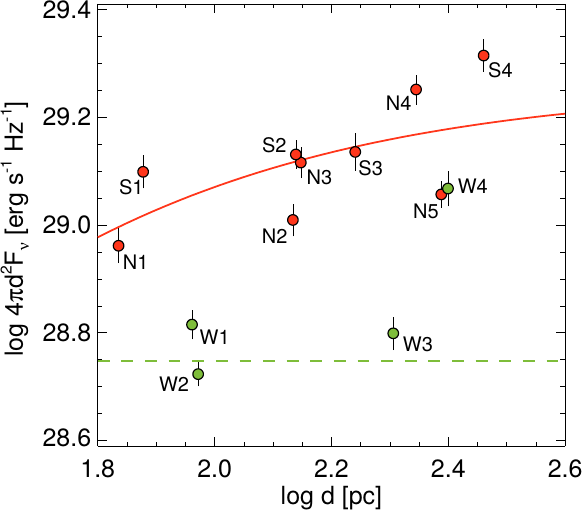}
\caption{Logarithm of the product of the squared distance of each region to the nucleus, $d^2$, and the 4.7\micron\ flux, $F_\nu$, required to produce CO $v$=1--0 band emission detected in each region according to the MC RT model (Sect.~\ref{ss:fit}) vs. $d$. The red circles correspond to regions located along the outflow axis, and the green circles correspond to off-axis regions (see Fig~\ref{fig_line_maps}).
The dashed green line indicates the $d^2F_\nu$ derived from the observed nuclear spectrum (i.e., edge-on luminosity of the obscuring envelope) assuming isotropic emission ($F_\nu= L_\nu \slash (4\pi d^2)$). The solid red line is the best-fit model assuming that a collimated beam illuminates the outflow (Eq.~\ref{eq:coll})
\label{fig_lumdist}}
\end{figure}

Bipolar outflows in AGNs are typically launched along directions perpendicular to the obscuring torus of gas and dust (e.g., \citealt{Honig2017}). The outflow of NGC~3256\,S lies very close to the plane of the sky, so it is likely that the torus of NGC~3256\,S is seen nearly edge-on.
\citet{Efstathiou2022} showed that the IR emission from the torus in IR-bright galaxies can be highly anisotropic, with corrections increasing the inferred AGN luminosity by up to a factor of $\sim$10 for nearly edge-on tori.
Therefore, the anisotropic 4.7\micron\ near-IR radiation field suggested by the MC RT modeling in Sect.~\ref{ss:fit} can be consistent with the presence of an edge-on torus. 

This edge-on torus can also account for the fact that in the three regions W1, W2, and W3, located far from the outflow axis and thus viewing the torus less strongly face-on, the near-IR radiation field agrees with that observed along our line of sight.
We note, however, that for region W4, which is also expected to view the torus edge-on based on its projected location, the required radiation field is approximately a factor of two higher than expected. This might indicate that the obscuring material in the torus is clumpy, allowing the near-IR radiation to escape along some directions other than the polar axis.

In addition to the anisotropy of the nuclear near-IR radiation field, the MC RT models indicate that its intensity does not follow a $d^{-2}$ scaling, with higher intensities required at larger distances relative to the $d^{-2}$ case (from about two times brighter at 70\,pc to about five times brighter at 300\,pc). {The intensities significantly deviate from the $d^{-2}$ case, as indicated by a Spearman rank correlation coefficient of 0.7 and a p value of 0.036. (see Fig.~\ref{fig_lumdist}).}
This behavior can be explained if the radiation from the AGN is collimated into a relatively narrow solid angle. Alternatively, a fast decrease in the AGN luminosity by a factor of 2 over the last $\sim$1000\,yr (corresponding to the light-travel time between N1 and N5) might also explain the deviations from the $d^{-2}$ relation.
{However, a luminosity decline would act isotropically and should therefore also enhance the off-axis regions, whereas W1--W3  are consistent with the line-of-sight value.
}

In this case, the hot dust heated by the AGN emitting in the near-IR is initially located inside an optically thick envelope of colder dust (e.g., \citealt{GonzalezAlfonso2019}). A fast outflow is launched by the AGN, potentially associated with the radio jet suggested by the extended 3.6\,cm radio continuum of NGC~3256\,S (see \citealt{Neff2003, Sakamoto2014}) {and escapes the dust cocoon through a channel that collimates the near-IR emission from the inner hotter dust core photosphere.}
Under these conditions, the collimated near-IR flux density, $F_\nu$, seen by the molecular clouds in the outflow { base} is $L_\nu^{\rm beam}$\slash $A(d)$, where $L_\nu^{\rm beam}$ is the spectral luminosity density emitted within the conical beam, and $A(d)$ is the area of the beam at a distance $d$ from the surface of the envelope.
For a collimated beam with a small half-opening angle $\alpha$ and radius $r_0$ at the surface of the dust envelope, the distance to the virtual apex of the cone is $r_0\slash \tan \alpha$. Hence, $A(d)=\Omega_\alpha (r_0\slash \tan \alpha + d)^2$, where $\Omega_\alpha$ is the solid angle of the beam, and

\begin{equation}
F_\nu = \frac{L_\nu^{\rm beam}}{\Omega_\alpha  \left[ \frac{r_0}{\tan \alpha} + d \right]^2} = \frac{L_\nu^{\rm core}}{4 \pi \left[ \frac{r_0}{\tan \alpha} + d \right]^2}
\label{eq:coll}
,\end{equation}
where $L_\nu^{\rm beam}$ is related to the total spectral luminosity density of the core through $L_\nu^{\rm beam} = L_\nu^{\rm core} \frac{\Omega_\alpha}{4\pi}$, assuming that the core emits isotropically.

Fig.~\ref{fig_lumdist} shows that except W4, the off-axis regions (green circles) agree with the expected illumination due to the edge-on torus with $L_\nu$ = 10$^{28.75}$\,erg\,s$^{-1}$\,Hz$^{-1}$ at 4.7\micron. However, all the regions along the outflow axis (red circles) have higher fluxes that can be reproduced by the collimated beam model.
The parameters of the best-fit model are $\frac{r_0}{\tan \alpha}$ = 24$\pm$9\,pc and $L_\nu^{\rm core}$ = 10$^{29.25 \pm 0.06}$\,erg\,s$^{-1}$\,Hz$^{-1}$, which is about three times larger than the 4.7\micron\ $L_\nu$ measured along our edge-on line of sight. 
This can be explained by a higher dust temperature, $T_d$, of the inner core (15--25\% greater than that of the outer envelope for $T_d$=300--600\,K), a deviation of the intrinsic core emission from isotropy, or by a combination of the two.

To estimate $r_0$, we approximated the beam half-opening angle from the width of the molecular outflow, $w$, at a distance $d$, using the relation
\begin{equation}
\tan \alpha = \frac{w}{2 \left(\frac{r_0}{\tan \alpha} + d\right)},
\end{equation}
where $\frac{r_0}{\tan \alpha}$ was obtained from the fit. We found $\alpha\sim$15--20\degree (see Fig.~\ref{fig_line_maps}), which implies that the radius of the opening in the dust envelope is $r_0$$\sim$6--9\,pc. The diameter of this opening is slightly smaller than but comparable to the envelope diameter estimated from the deconvolved FWHM of the 4.7\micron\ continuum (0\farcs12; $\sim$25\,pc) in the NIRSpec data\footnote{We used the calibration star TYC\,4433-1800-1 to estimate the point spread function FWHM, which was then subtracted in quadrature from the measured FWHM of the NGC~3256\,S nucleus.}.

\begin{figure}
\centering
\includegraphics[width=0.33\textwidth]{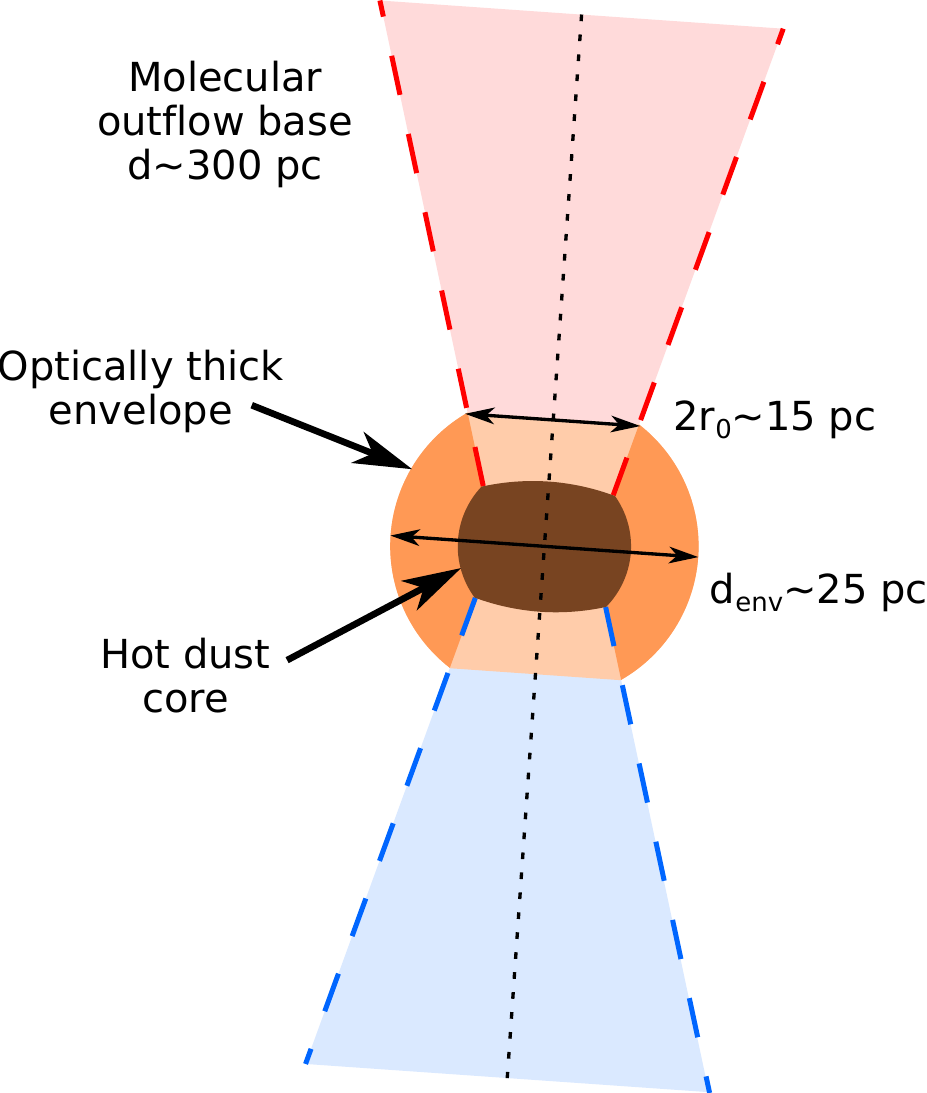}
\caption{Schematic diagram representing the central region { up to $\sim$300\,pc from the core} of NGC~3256\,S. The hot dust core (dark brown) heated by the AGN is surrounded by an optically thick envelope of colder material (orange).
A channel with a half-opening angle $\alpha$ of $\sim$15--20\degree\ transverses the envelope and collimates the hot core 4.7\micron\ near-IR radiation that illuminates the molecular gas { at the outflow base} (blue and red shaded areas corresponding to the blue- and redshifted outflow lobes, respectively) and produces the observed CO $v$=1--0 band. The dotted black line indicates the outflow axis. { The schematic is not to scale.}
\label{fig_torus}}
\end{figure}

\subsection{A low-luminosity AGN in an expanding optically and geometrically thick envelope}\label{ss:agn}

Figure~\ref{fig_torus} shows a schematic of the nuclear region of NGC~3256\,S, including the structures needed to explain the properties of the CO $v$=1--0 band and the nuclear near-IR continuum.
This object hosts a heavily obscured AGN with a relatively modest IR luminosity, (1--5)$\times$10$^9$\,\Lsun\ \citep{Ohyama2015}. Based on our modeling, this obscured AGN is the main power source of the hot dust core, $\nu L_\nu^{\rm core}\sim$4$\times$10$^9$\,\Lsun, which is further reprocessed by an optically thick dusty envelope.

No high-ionization emission lines such as [\ion{Ne}{vi}]7.6\micron\ or [\ion{Ne}{v}]14.3\micron\ (with ionization potentials $>$95\,eV) typically associated with AGNs are detected in the mid-IR spectrum of NGC~3256\,S (e.g., \citealt{Pereira2010, Bohn2024, GarciaBernete2025}).
This is similar to the case of other deeply obscured AGNs in local (U)LIRGs, which likewise lack high-ionization lines in their spectra (e.g., \citealt{Farrah07, GarciaBernete2024_SODA, Perna2024, Buiten2024, Goldberg2024, AAH2024, Rieke2025}), where AGNs are identified through extreme luminosity surface densities (e.g., \citealt{BarcosMunoz2017, Pereira2021, GonzalezAlfonso2026}), excited molecular ro-vibrational bands (e.g., \citealt{GarciaBernete2024_SODA}), and anomalous PAH band ratios and equivalent widths (e.g., \citealt{GarciaBernete2022_CON, Rich2023, Donnan2023}).
Therefore, the inner hot dust core of {NGC~3256\,S} must be optically thick in the mid-IR (with H column density $N_{\rm H}$>10$^{23}$\,cm$^{-2}$) and geometrically thick with a high covering factor, preventing the escape of ionizing radiation from the AGN and the formation of an extended highly ionized narrow-line region as in Seyfert-like AGNs.

Extremely obscured optically thick nuclear cores with high covering factors are associated with CONs in (U)LIRGs, which can reach $N_{\rm H}$$>$10$^{25}$\,cm$^{-2}$ (e.g., \citealt{GonzalezAlfonso2019}). 
In CONs, the trapping of continuum photons due to high dust opacities produces relatively cold SED peaking at 20--40\micron, with outer envelope dust temperatures of $T_d$=80--200\,K, even if $T_d$ in the inner layers is much higher due to the {greenhouse effect} (up to 1000\,K; \citealt{GonzalezAlfonso2019, GonzalezAlfonso2026}).
In contrast, the SED of NGC~3256\,S peaks at shorter wavelengths ($\sim$5--10\micron; \citealt{Ohyama2015}), suggesting a hotter outer envelope, and therefore, an $N_{\rm H}$$<$10$^{24}$\,cm$^{-2}$ lower than that of CONs, as well as less reprocessing of continuum photons.
A direct consequence of this hotter outer envelope is that the rotational HCN-vib lines in the submillimeter, commonly observed in CONs, are not detected in NGC~3256\,S \citep{Falstad2021}, since its 14\micron\ continuum, which excites the HCN $v_2$ vibrational band and leads to the submillimeter HCN-vib emission, is relatively weak.

To further constrain $N_{\rm H}$, we used the H$_2$O 3.02\micron\ ice absorption. This ice absorption is very strong in the nuclear continuum of NGC~3256\,S and shows a saturated profile from which we derived $N_{\rm H}$$\sim$10$^{23}$\,cm$^{-2}$ (\citealt{Rigopoulou2024, Speranza26}).
This deep ice absorption likely arises from a combination of the prominent dust lane crossing the nucleus of NGC~3256\,S (see \citealt{Lira2002, Zepf1999}) and colder dust in the outer regions of the nuclear envelope, similar to the material producing the 6\micron\ water-ice absorption associated with the outer layers of the torus in type 2 Seyfert AGNs \citep{GarciaBernete2024_ice}.

{The molecular outflow seems to be escaping through a channel in the optically thick envelope (Sect.~\ref{ss:anisotropy}).
However, the origin of this channel remains unclear. It may have been created by the detected radio jet \citep{Neff2003, Sakamoto2014}, carved by the molecular outflow itself, or formed through a combination of the two processes. Alternatively, the channel might correspond to a preexisting structure established by the accretion flow.
In addition, } the envelope itself appears to be expanding, as indicated by the nuclear blueshifted high-excitation CO $v$=1--0 absorption relative to the low-excitation CO $v$=1--0 absorption that lies at the CO(2--1) systemic velocity of NGC~3256\,S \citepalias{Pereira2024CO}.
Therefore, these results suggest that the AGN of NGC~3256\,S is beginning to clear its surrounding environment as a low-luminosity AGN after building up a column density of $N_{\rm H}$$\sim$10$^{22-23}$\,cm$^{-2}$, which is lower than that of a CON, but comparable to that of typical Seyfert 2 AGN \citep{Ricci2017_BAT}. However, a high covering factor, likely sustained by stochastic inflows during the galaxy interaction, prevents the escape of ionizing photons from the AGN, similar to a CON, but in contrast to typical Seyfert 2 AGNs.

\subsection{Core luminosity from the CO emission}\label{ss:band emission}

We related the obscured core luminosity, $L_\nu^{\rm core}$, to the CO $v$=1--0 band luminosity. From Eq.~\ref{eq_lagn}, we defined a luminosity-weighted factor, $k_{\rm p}$, that encapsulates the dependence on the physical conditions of the molecular gas ($T_{\rm kin}$, $n_{\rm H_2}$, and $N_{\rm CO}^{\rm max}\slash \Delta{\rm v}$).
We defined $k_{\rm p}=2\slash 3 \times N_{\rm CO}^{\rm max}\slash \Delta{\rm v} \times \sum{r_i L_i}\slash L_{{\rm band}}$, where $L_{{\rm band}}$ is the total band luminosity $\sum L_{{\rm line},i}$. Then, Eq.~\ref{eq_lagn} can be expressed as

\begin{equation}
F_\nu(d) = k_p \frac{L_{\rm line, i}}{\mathcal{N}} \label{eq_average}.
\end{equation}

\noindent The $\mathcal{N}$ can be approximated from the luminosity of the \hbox{CO(1--0)} rotational line, $L_{\rm CO(J=1-0)}$, using the scaling factor, $\alpha_{\rm CO} X_{\rm CO}$, where $X_{\rm CO}$ is the CO abundance. We found that this scaling factor varies by a factor of $\sim$2 on average across the NGC~3256\,S outflow { base} \citepalias{Pereira2024CO}.
Finally, taking into account the geometrical parameters of the collimated beam from the core ($\alpha$ and $r_0$ in Eq.~\ref{eq:coll}) and combining Eqs.~\ref{eq:coll} and \ref{eq_average}, we obtained

\begin{equation}
L_\nu^{\rm core} = k \frac{L_{\rm band}}{L_{\rm CO(J=1-0)}},
\end{equation}

\noindent where $k$ is a constant that depends on the spatial distribution and average physical conditions of the molecular gas in the outflow { base}.

{
Using the observed CO luminosities {in the outflow apertures (Table~\ref{tbl_bestfit}) and the derived $L_\nu^{\rm core}$ (Sect.~\ref{ss:anisotropy})}, we determined individual values of $k$ for each region. Since the illuminating flux decreases as ($r_0\slash \tan(\alpha) + d)^{-2}$ (Eq.~\ref{eq:coll}), $k$ increases with distance, from $\sim$1$\times$10$^{-3}$ (N1, $d$=68\,pc) to $\sim$8$\times$10$^{-3}$\,K\,km\,s$^{-1}${pc$^2$}\,Hz$^{-1}$ (N5, $d$=244\,pc).
A representative value of $k$ for the central $\sim$300\,pc probed by the NIRSpec data was obtained from the median of the nine on-axis apertures (N1--N5 and S1--S4), $k$=(3.5 $\pm$ 1.1)$\times$10$^{-3}$\,K\,km\,s$^{-1}${pc$^2$}\,Hz$^{-1}$,} which can be used to estimate the core near-IR luminosity of similar objects that host deeply obscured AGNs and analogous collimated molecular outflows, as can be the case of some local LIRGs and ULIRGs.

\section{Conclusions}\label{s:conclusions}
 
The southern nucleus of the interacting luminous IR system NGC~3256 ($d$=40\,Mpc; $L_{\rm IR}$=4$\times$10$^{11}$\,\Lsun) hosts a deeply obscured low-luminosity AGN ($L_{\rm 2-10\,keV}\sim10^{40}$\,erg\,s$^{-1}$) and a massive bipolar collimated molecular outflow ({ mass} outflow rate $>50$\,\Msun\,yr$^{-1}$), oriented nearly in the plane of the sky (e.g., \citealt{Sakamoto2014, Emonts2014, Ohyama2015}).
We used \JWST\slash NIRSpec data to study the spatially resolved CO $v$=1--0 4.67\micron\ ro-vibrational band emission mostly detected along the {base of the} bipolar molecular outflow.
We extended the analysis of this band presented in \citetalias{Pereira2024CO} by using MC RT models and a non-LTE treatment of the CO excitation. With this new models, we investigated the angular anisotropy of the near-IR radiation field produced by the obscured AGN core that excites the CO $v$=1–0 band. The main results from this work are listed below.

\begin{enumerate}

\item \textit{Anisotropic radiation field}. Using a grid of non-LTE MC RT models, we successfully fitted the fluxes of all the ro-vibrational lines of the CO $v$=1--0 band in 13 apertures {($r$=0\farcs2; $\sim$40\,pc)} where this band is detected in emission. From this fit, the $\sim$4.7\micron\ spectral luminosity density, $L_\nu$, from the NGC~3256\,S core, as seen by the material {at the base of the} bipolar outflow, is approximately two to four times higher than that inferred from the core spectrum along our line of sight. This indicates that the near-IR radiation field in this object is anisotropic.

\item {\it Collimated radiation field.} In addition, the radiation field exciting the CO $v$=1--0 band {at the base of the} outflow does not decrease with distance according to the expected $d^{-2}$ law. Instead, the observed intensity scaling can be explained if the radiation from the core is collimated into a relatively narrow solid angle. Assuming a collimated beam, with a half-opening angle $\alpha$ of 15--20\degree\ based on the extent of the bipolar molecular outflow { base}, we found that the diameter of the structure collimating the radiation is 15$\pm$5\,pc. This is slightly smaller than but comparable to the diameter of the obscuring envelope ($\sim$25$\pm$5\,pc).
The estimated about three times higher $L_\nu$ at 4.7\micron\ toward the outflow { base} indicates that it is either illuminated by dust hotter than that observed along our line of sight or that the intrinsic core emission is anisotropic at smaller scales.

\item {\it Expanding optically and geometrically thick envelope.} To date, no high-ionization lines have been detected in NGC~3256\,S, including in its mid-IR spectrum. This indicates that the AGN is surrounded by optically thick material even in the mid-IR ($N_{\rm H}$>10$^{23}$\,cm$^{-2}$) with a high covering factor. The bipolar outflow and the increased IR radiation field along the outflow direction suggest that the geometry of the obscuring material resembles a torus that is seen almost edge-on in our line of sight. The bipolar outflow and the blueshifted CO $v$=1--0 absorption toward the nucleus suggests that the obscuring envelope is currently expanding.

\item {\it Hybrid CON\slash Seyfert 2 nature.} The SED of NGC~3256\,S peaks at shorter wavelengths ($\sim$5--10\micron) than CONs with high $N_{\rm H}$$>$10$^{25}$\,cm$^{-2}$, indicating a hotter dust photosphere in NGC~3256\,S. This implies a lower $N_{\rm H}$ than CONs. 
The estimated $N_{\rm H}$ range of $\sim$10$^{22-23}$\,cm$^{-2}$ for NGC~3256\,S is comparable to that of Seyfert 2 AGNs. However, the absence of high-ionization lines indicates a high covering factor of the obscuring material, similar to that expected in CONs and in contrast to Seyfert 2. This high covering factor might be sustained by inflows during this galactic interaction. Therefore, NGC~3256\,S shares properties with CONs and Seyfert 2 AGN.

\item {\it Core IR luminosity versus CO luminosity relation.} The CO $v$=\hbox{1--0} band emission, $L_{\rm band}$, is proportional to the 4.7\micron\ $L_\nu$ from the core and the molecular gas mass. Thus, by combining the CO($J$=1--0) luminosity, which traces the molecular gas mass and the estimated 4.7\micron\ $L_\nu$ illuminating the outflow { base}, we found that for NGC~3256\,S, $L_\nu = k {L_{\rm band}}\slash {L_{\rm CO(J=1-0)}}$ with { a median value of} $k$ =(3.5 $\pm$ 1.1)$\times$10$^{-3}$\,{ K\,km\,s$^{-1}${pc$^2$}\,Hz$^{-1}$ for the studied regions}. This value of $k$, which depends on the physical conditions of the molecular gas and the outflow geometrical parameters, is representative of the { base of the} outflow of NGC~3256\,S. Therefore, this approach might be used to estimate $L_\nu$ only in other obscured AGN hosting similar bipolar collimated molecular outflows as some local (U)LIRGs.

\end{enumerate}

\begin{acknowledgements}
{We thank the referee for a careful reading of the manuscript and useful suggestions.}
The authors acknowledge the GOALS team for developing their observing programs.
MPS acknowledges support from grants RYC2021-033094-I, CNS2023-145506, and PID2023-146667NB-I00 funded by MCIN/AEI/10.13039/501100011033 and the European Union NextGenerationEU/PRTR.

This work is based on observations made with the NASA/ESA/CSA James Webb Space Telescope. The data were obtained from the Mikulski Archive for Space Telescopes at the Space Telescope Science Institute, which is operated by the Association of Universities for Research in Astronomy, Inc., under NASA contract NAS 5-03127 for JWST; and from the European JWST archive (eJWST) operated by the ESAC Science Data Centre (ESDC) of the European Space Agency. These observations are associated with program \#1328.
This paper makes use of the following ALMA data: ADS/JAO.ALMA\#2018.1.00223.S.
ALMA is a partnership of ESO (representing its member states), NSF (USA) and NINS (Japan), together with NRC (Canada) and NSC and ASIAA (Taiwan) and KASI (Republic of Korea), in cooperation with the Republic of Chile. The Joint ALMA Observatory is operated by ESO, AUI/NRAO and NAOJ.
The National Radio Astronomy Observatory is a facility of the National Science Foundation operated under cooperative agreement by Associated Universities, Inc.

\end{acknowledgements}

\appendix

\onecolumn
\section{MC RT fitting results for all regions}\label{apx:figs}
\begin{figure}[h]
\centering
\includegraphics[width=0.97\textwidth]{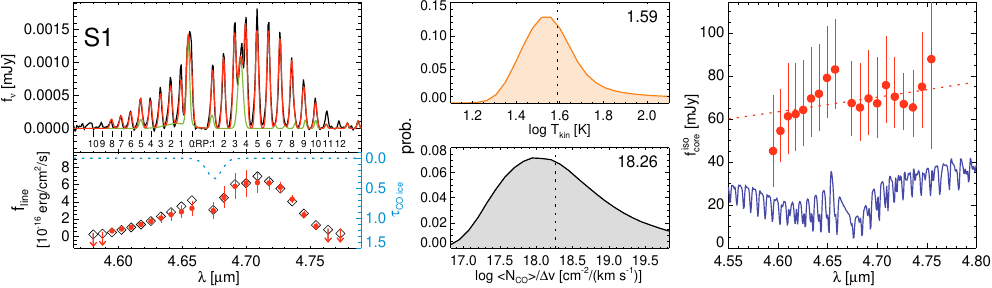}
\includegraphics[width=0.97\textwidth]{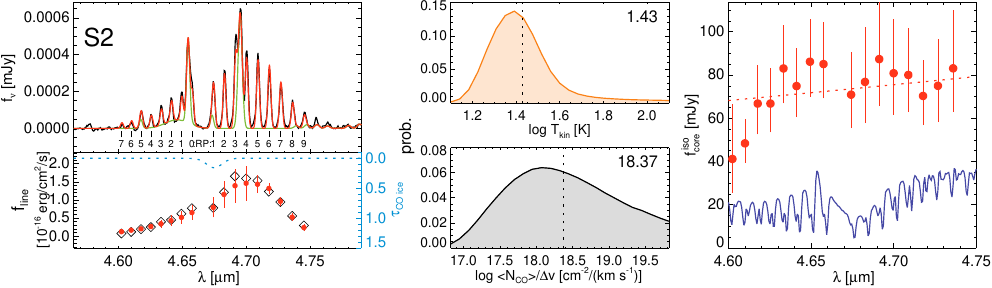}
\includegraphics[width=0.97\textwidth]{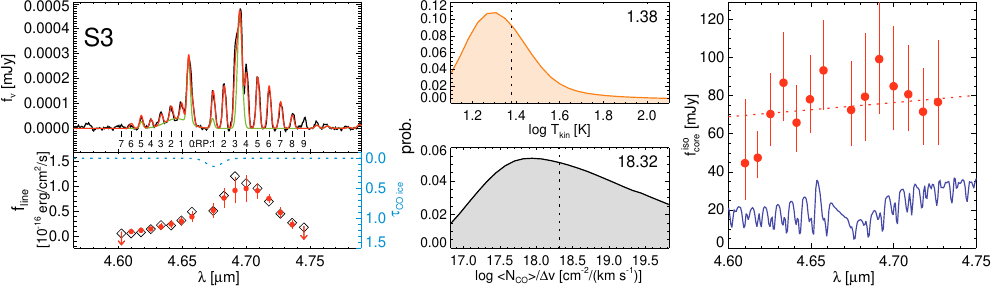}
\includegraphics[width=0.97\textwidth]{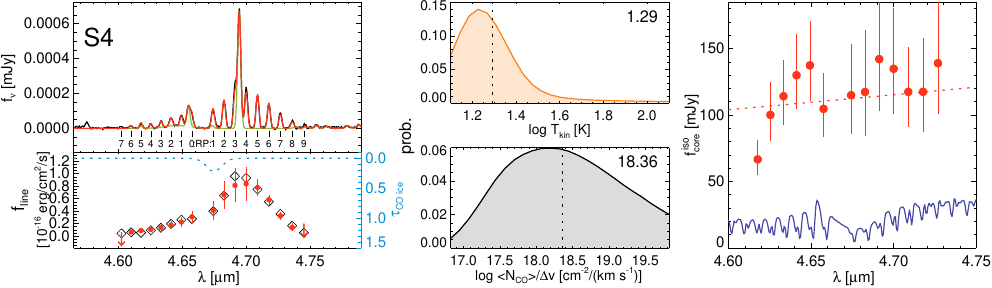}
\caption{Continuation of Fig.~\ref{fig_bestfit}.\label{fig_bestfit_apx}}
\end{figure}

\addtocounter{figure}{-1} %
\begin{figure}
\centering
\includegraphics[width=0.97\textwidth]{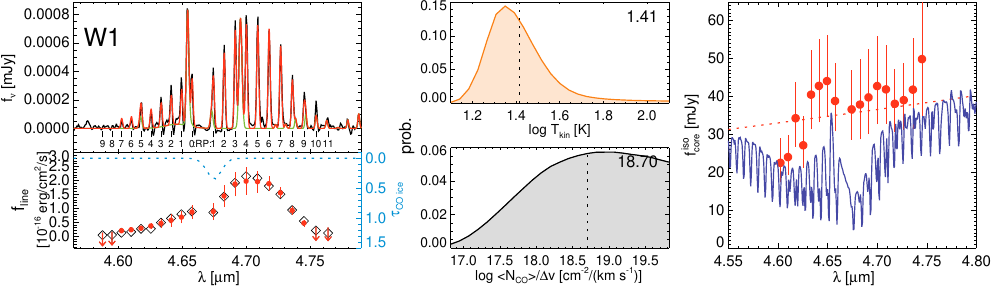}
\includegraphics[width=0.97\textwidth]{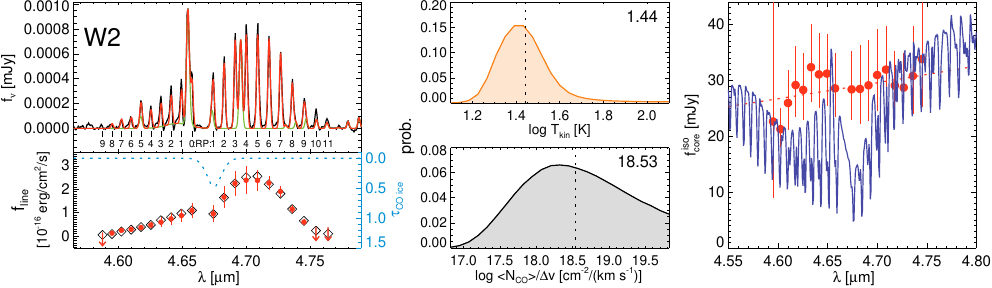}
\includegraphics[width=0.97\textwidth]{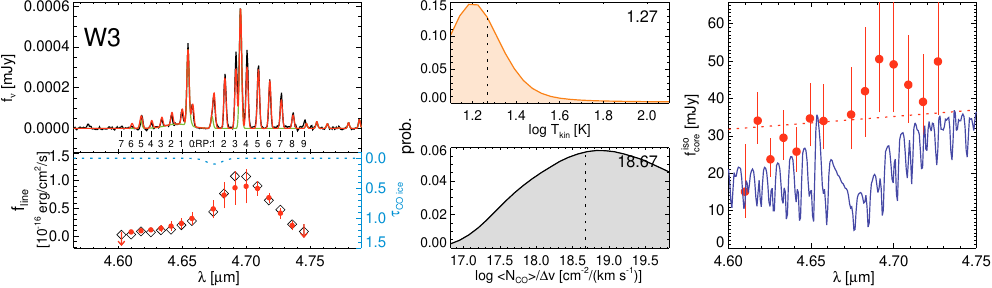}
\includegraphics[width=0.97\textwidth]{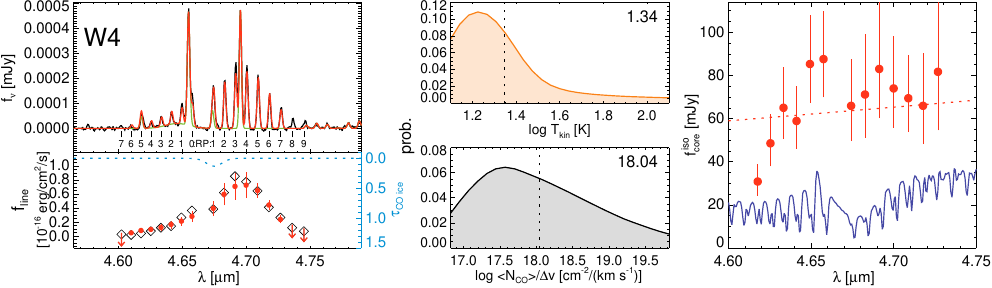}
\caption{Continued.}
\end{figure}

\addtocounter{figure}{-1} %
\begin{figure}
\centering
\includegraphics[width=0.97\textwidth]{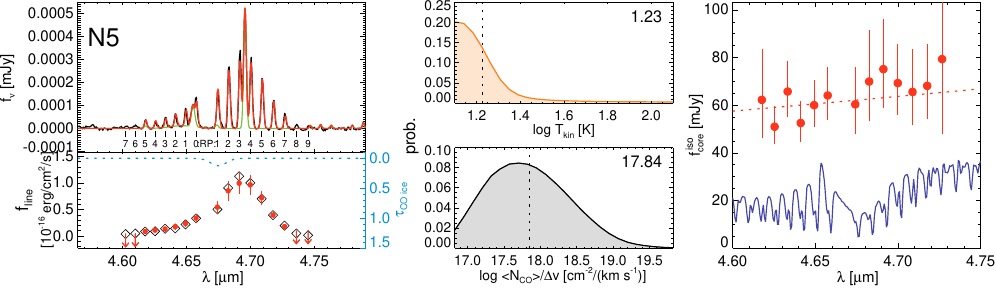}
\caption{Continued.}
\end{figure}


\begin{thebibliography}{60}
\expandafter\ifx\csname natexlab\endcsname\relax\def\natexlab#1{#1}\fi

\bibitem[{{Alonso Herrero} {et~al.}(2024){Alonso Herrero}, {Hermosa Mu{\~n}oz},
  {Labiano}, {Guillard}, {Buiten}, {Dicken}, {van der Werf},
  {{\'A}lvarez-M{\'a}rquez}, {B{\"o}ker}, {Colina}, {Eckart},
  {Garc{\'\i}a-Mar{\'\i}n}, {Jones}, {Pantoni}, {P{\'e}rez-Gonz{\'a}lez},
  {Rouan}, {Ward}, {Baes}, {{\"O}stlin}, {Royer}, {Wright}, {G{\"u}del},
  {Henning}, {Lagage}, \& {van Dishoeck}}]{AAH2024}
{Alonso Herrero}, A., {Hermosa Mu{\~n}oz}, L., {Labiano}, A., {et~al.} 2024,
  \aap, 690, A95

\bibitem[{{Alonso-Herrero} {et~al.}(2012){Alonso-Herrero}, {Pereira-Santaella},
  {Rieke}, \& {Rigopoulou}}]{AAH2012a}
{Alonso-Herrero}, A., {Pereira-Santaella}, M., {Rieke}, G.~H., \& {Rigopoulou},
  D. 2012, \apj, 744, 2

\bibitem[{{Barcos-Mu{\~n}oz} {et~al.}(2018){Barcos-Mu{\~n}oz}, {Aalto},
  {Thompson}, {Sakamoto}, {Mart{\'\i}n}, {Leroy}, {Privon}, {Evans}, \&
  {Kepley}}]{BarcosMunoz2018}
{Barcos-Mu{\~n}oz}, L., {Aalto}, S., {Thompson}, T.~A., {et~al.} 2018, \apjl,
  853, L28

\bibitem[{{Barcos-Mu{\~n}oz} {et~al.}(2017){Barcos-Mu{\~n}oz}, {Leroy},
  {Evans}, {Condon}, {Privon}, {Thompson}, {Armus}, {D{\'\i}az-Santos},
  {Mazzarella}, {Meier}, {Momjian}, {Murphy}, {Ott}, {Sanders}, {Schinnerer},
  {Stierwalt}, {Surace}, \& {Walter}}]{BarcosMunoz2017}
{Barcos-Mu{\~n}oz}, L., {Leroy}, A.~K., {Evans}, A.~S., {et~al.} 2017, \apj,
  843, 117

\bibitem[{{Bohn} {et~al.}(2024){Bohn}, {Inami}, {Togi}, {Armus}, {Lai},
  {Barcos-Munoz}, {Song}, {Linden}, {Surace}, {Bianchin}, {U}, {Evans},
  {B{\"o}ker}, {Malkan}, {Larson}, {Stierwalt}, {Buiten}, {Charmandaris},
  {Diaz-Santos}, {Howell}, {Privon}, {Ricci}, {van der Werf}, {Aalto},
  {Hayward}, {Kader}, {Mazzarella}, {Muller-Sanchez}, \& {Sanders}}]{Bohn2024}
{Bohn}, T., {Inami}, H., {Togi}, A., {et~al.} 2024, \apj, 977, 36

\bibitem[{{Boogert} {et~al.}(2015){Boogert}, {Gerakines}, \&
  {Whittet}}]{Boogert2015}
{Boogert}, A.~C.~A., {Gerakines}, P.~A., \& {Whittet}, D. C.~B. 2015, \araa,
  53, 541

\bibitem[{{Buiten} {et~al.}(2024){Buiten}, {van der Werf}, {Viti}, {Armus},
  {Barr}, {Barcos-Mu{\~n}oz}, {Evans}, {Inami}, {Linden}, {Privon}, {Song},
  {Rich}, {Aalto}, {Appleton}, {B{\"o}ker}, {Charmandaris}, {Diaz-Santos},
  {Hayward}, {Lai}, {Medling}, {Ricci}, \& {U}}]{Buiten2024}
{Buiten}, V.~A., {van der Werf}, P.~P., {Viti}, S., {et~al.} 2024, \apj, 966,
  166

\bibitem[{{Cicone} {et~al.}(2014){Cicone}, {Maiolino}, {Sturm},
  {Graci{\'a}-Carpio}, {Feruglio}, {Neri}, {Aalto}, {Davies}, {Fiore},
  {Fischer}, {Garc{\'{\i}}a-Burillo}, {Gonz{\'a}lez-Alfonso},
  {Hailey-Dunsheath}, {Piconcelli}, \& {Veilleux}}]{Cicone2014}
{Cicone}, C., {Maiolino}, R., {Sturm}, E., {et~al.} 2014, \aap, 562, A21

\bibitem[{{D{\'{\i}}az-Santos} {et~al.}(2010){D{\'{\i}}az-Santos},
  {Alonso-Herrero}, {Colina}, {Packham}, {Levenson}, {Pereira-Santaella},
  {Roche}, \& {Telesco}}]{DiazSantos2010}
{D{\'{\i}}az-Santos}, T., {Alonso-Herrero}, A., {Colina}, L., {et~al.} 2010,
  \apj, 711, 328

\bibitem[{{Donnan} {et~al.}(2023){Donnan}, {Garc{\'\i}a-Bernete}, {Rigopoulou},
  {Pereira-Santaella}, {Alonso-Herrero}, {Roche}, {Hern{\'a}n-Caballero}, \&
  {Spoon}}]{Donnan2023}
{Donnan}, F.~R., {Garc{\'\i}a-Bernete}, I., {Rigopoulou}, D., {et~al.} 2023,
  \mnras, 519, 3691

\bibitem[{{Efstathiou} {et~al.}(2022){Efstathiou}, {Farrah}, {Afonso},
  {Clements}, {Gonz{\'a}lez-Alfonso}, {Lacy}, {Oliver}, {Papadopoulou Lesta},
  {Pearson}, {Rigopoulou}, {Rowan-Robinson}, {Spoon}, {Verma}, \&
  {Wang}}]{Efstathiou2022}
{Efstathiou}, A., {Farrah}, D., {Afonso}, J., {et~al.} 2022, \mnras, 512, 5183

\bibitem[{{Emonts} {et~al.}(2014){Emonts}, {Piqueras-L{\'o}pez}, {Colina},
  {Arribas}, {Villar-Mart{\'{\i}}n}, {Pereira-Santaella}, {Garcia-Burillo}, \&
  {Alonso-Herrero}}]{Emonts2014}
{Emonts}, B.~H.~C., {Piqueras-L{\'o}pez}, J., {Colina}, L., {et~al.} 2014,
  \aap, 572, A40

\bibitem[{{Falstad} {et~al.}(2021){Falstad}, {Aalto}, {K{\"o}nig}, {Onishi},
  {Muller}, {Gorski}, {Sato}, {Stanley}, {Combes}, {Gonz{\'a}lez-Alfonso},
  {Mangum}, {Evans}, {Barcos-Mu{\~n}oz}, {Privon}, {Linden},
  {D{\'\i}az-Santos}, {Mart{\'\i}n}, {Sakamoto}, {Harada}, {Fuller},
  {Gallagher}, {van der Werf}, {Viti}, {Greve}, {Garc{\'\i}a-Burillo},
  {Henkel}, {Imanishi}, {Izumi}, {Nishimura}, {Ricci}, \&
  {M{\"u}hle}}]{Falstad2021}
{Falstad}, N., {Aalto}, S., {K{\"o}nig}, S., {et~al.} 2021, \aap, 649, A105

\bibitem[{{Farrah} {et~al.}(2007){Farrah}, {Bernard-Salas}, {Spoon}, {Soifer},
  {Armus}, {Brandl}, {Charmandaris}, {Desai}, {Higdon}, {Devost}, \&
  {Houck}}]{Farrah07}
{Farrah}, D., {Bernard-Salas}, J., {Spoon}, H.~W.~W., {et~al.} 2007, \apj, 667,
  149

\bibitem[{{Garc{\'\i}a-Bernete}
  {et~al.}(2024{\natexlab{a}}){Garc{\'\i}a-Bernete}, {Alonso-Herrero},
  {Rigopoulou}, {Pereira-Santaella}, {Shimizu}, {Davies}, {Donnan}, {Roche},
  {Gonz{\'a}lez-Mart{\'\i}n}, {Ramos Almeida}, {Bellocchi}, {Boorman},
  {Combes}, {Efstathiou}, {Esparza-Arredondo}, {Garc{\'\i}a-Burillo},
  {Gonz{\'a}lez-Alfonso}, {Hicks}, {H{\"o}nig}, {Labiano}, {Levenson},
  {L{\'o}pez-Rodr{\'\i}guez}, {Ricci}, {Packham}, {Rouan}, {Stalevski}, \&
  {Ward}}]{GarciaBernete2024_ice}
{Garc{\'\i}a-Bernete}, I., {Alonso-Herrero}, A., {Rigopoulou}, D., {et~al.}
  2024{\natexlab{a}}, \aap, 681, L7

\bibitem[{{Garc{\'\i}a-Bernete} {et~al.}(2025){Garc{\'\i}a-Bernete}, {Donnan},
  {Rigopoulou}, {Pereira-Santaella}, {Gonz{\'a}lez-Alfonso}, {Thatte}, {Aalto},
  {K{\"o}nig}, {Maksymowicz-Maciata}, {Smith}, {Huang}, {Magdis}, {Roche},
  {Devriendt}, \& {Slyz}}]{GarciaBernete2025}
{Garc{\'\i}a-Bernete}, I., {Donnan}, F.~R., {Rigopoulou}, D., {et~al.} 2025,
  \aap, 696, A135

\bibitem[{{Garc{\'\i}a-Bernete}
  {et~al.}(2024{\natexlab{b}}){Garc{\'\i}a-Bernete}, {Pereira-Santaella},
  {Gonz{\'a}lez-Alfonso}, {Rigopoulou}, {Efstathiou}, {Donnan}, \&
  {Thatte}}]{GarciaBernete2024_SODA}
{Garc{\'\i}a-Bernete}, I., {Pereira-Santaella}, M., {Gonz{\'a}lez-Alfonso}, E.,
  {et~al.} 2024{\natexlab{b}}, \aap, 682, L5

\bibitem[{{Garc{\'\i}a-Bernete} {et~al.}(2022){Garc{\'\i}a-Bernete},
  {Rigopoulou}, {Aalto}, {Spoon}, {Hern{\'a}n-Caballero}, {Efstathiou},
  {Roche}, \& {K{\"o}nig}}]{GarciaBernete2022_CON}
{Garc{\'\i}a-Bernete}, I., {Rigopoulou}, D., {Aalto}, S., {et~al.} 2022, \aap,
  663, A46

\bibitem[{{Goldberg} {et~al.}(2024){Goldberg}, {Buiten}, {Rieke},
  {Alonso-Herrero}, {Paggi}, {van der Werf}, {Stone}, {Morrison}, {Alberts},
  {Dicken}, \& {Wright}}]{Goldberg2024}
{Goldberg}, C.~E., {Buiten}, V.~A., {Rieke}, G.~H., {et~al.} 2024, \apj, 977,
  55

\bibitem[{{Gonz{\'a}lez-Alfonso} {et~al.}(1998){Gonz{\'a}lez-Alfonso},
  {Cernicharo}, {van Dishoeck}, {Wright}, \& {Heras}}]{GonzalezAlfonso1998}
{Gonz{\'a}lez-Alfonso}, E., {Cernicharo}, J., {van Dishoeck}, E.~F., {Wright},
  C.~M., \& {Heras}, A. 1998, \apjl, 502, L169

\bibitem[{{Gonz{\'a}lez-Alfonso} {et~al.}(2017){Gonz{\'a}lez-Alfonso},
  {Fischer}, {Spoon}, {Stewart}, {Ashby}, {Veilleux}, {Smith}, {Sturm},
  {Farrah}, {Falstad}, {Mel{\'e}ndez}, {Graci{\'a}-Carpio}, {Janssen}, \&
  {Lebouteiller}}]{GonzalezAlfonso2017}
{Gonz{\'a}lez-Alfonso}, E., {Fischer}, J., {Spoon}, H.~W.~W., {et~al.} 2017,
  \apj, 836, 11

\bibitem[{{Gonz{\'a}lez-Alfonso} {et~al.}(2026){Gonz{\'a}lez-Alfonso},
  {Pereira-Santaella}, {Garc{\'\i}a-Bernete}, {Fischer}, \&
  {Speranza}}]{GonzalezAlfonso2026}
{Gonz{\'a}lez-Alfonso}, E., {Pereira-Santaella}, M., {Garc{\'\i}a-Bernete}, I.,
  {Fischer}, J., \& {Speranza}, G. 2026, \aap, 706, A227

\bibitem[{{Gonz{\'a}lez-Alfonso} \& {Sakamoto}(2019)}]{GonzalezAlfonso2019}
{Gonz{\'a}lez-Alfonso}, E. \& {Sakamoto}, K. 2019, \apj, 882, 153

\bibitem[{{Gonz{\'a}lez-Alfonso} {et~al.}(2002){Gonz{\'a}lez-Alfonso},
  {Wright}, {Cernicharo}, {Rosenthal}, {Boonman}, \& {van
  Dishoeck}}]{GonzalezAlfonso2002}
{Gonz{\'a}lez-Alfonso}, E., {Wright}, C.~M., {Cernicharo}, J., {et~al.} 2002,
  \aap, 386, 1074

\bibitem[{{Goorvitch}(1994)}]{Goorvitch1994}
{Goorvitch}, D. 1994, \apjs, 95, 535

\bibitem[{{Gorski} {et~al.}(2024){Gorski}, {Aalto}, {K{\"o}nig}, {Wethers},
  {Yang}, {Muller}, {Onishi}, {Sato}, {Falstad}, {Mangum}, {Linden}, {Combes},
  {Mart{\'\i}n}, {Imanishi}, {Wada}, {Barcos-Mu{\~n}oz}, {Stanley},
  {Garc{\'\i}a-Burillo}, {van der Werf}, {Evans}, {Henkel}, {Viti}, {Harada},
  {D{\'\i}az-Santos}, {Gallagher}, \& {Gonz{\'a}lez-Alfonso}}]{Gorski2024}
{Gorski}, M.~D., {Aalto}, S., {K{\"o}nig}, S., {et~al.} 2024, \aap, 684, L11

\bibitem[{{H{\"o}nig} \& {Kishimoto}(2017)}]{Honig2017}
{H{\"o}nig}, S.~F. \& {Kishimoto}, M. 2017, \apjl, 838, L20

\bibitem[{{Hopkins} {et~al.}(2013){Hopkins}, {Cox}, {Hernquist}, {Narayanan},
  {Hayward}, \& {Murray}}]{Hopkins2013}
{Hopkins}, P.~F., {Cox}, T.~J., {Hernquist}, L., {et~al.} 2013, \mnras, 430,
  1901

\bibitem[{{Kawaguchi} {et~al.}(2020){Kawaguchi}, {Yutani}, \&
  {Wada}}]{Kawaguchi2020}
{Kawaguchi}, T., {Yutani}, N., \& {Wada}, K. 2020, \apj, 890, 125

\bibitem[{{Lamperti} {et~al.}(2022){Lamperti}, {Pereira-Santaella}, {Perna},
  {Colina}, {Arribas}, {Garc{\'\i}a-Burillo}, {Gonz{\'a}lez-Alfonso}, {Aalto},
  {Alonso-Herrero}, {Combes}, {Labiano}, {Piqueras-L{\'o}pez}, {Rigopoulou}, \&
  {van der Werf}}]{Lamperti2022}
{Lamperti}, I., {Pereira-Santaella}, M., {Perna}, M., {et~al.} 2022, \aap, 668,
  A45

\bibitem[{{Lankhaar} {et~al.}(2024){Lankhaar}, {Aalto}, {Wethers}, {Moldon},
  {Beswick}, {Gorski}, {K{\"o}nig}, {Yang}, {Mangum}, {Gallagher}, {Combes},
  {Rigopoulou}, {Gonz{\'a}lez-Alfonso}, {Muller}, {Garcia-Bernete}, {Henkel},
  {Nishimura}, \& {Ricci}}]{Lankhaar2024}
{Lankhaar}, B., {Aalto}, S., {Wethers}, C., {et~al.} 2024, \aap, 689, A163

\bibitem[{{Lira} {et~al.}(2008){Lira}, {Gonzalez-Corvalan}, {Ward}, \&
  {Hoyer}}]{Lira2008}
{Lira}, P., {Gonzalez-Corvalan}, V., {Ward}, M., \& {Hoyer}, S. 2008, \mnras,
  384, 316

\bibitem[{{Lira} {et~al.}(2002){Lira}, {Ward}, {Zezas}, {Alonso-Herrero}, \&
  {Ueno}}]{Lira2002}
{Lira}, P., {Ward}, M., {Zezas}, A., {Alonso-Herrero}, A., \& {Ueno}, S. 2002,
  \mnras, 330, 259

\bibitem[{{Lutz} {et~al.}(2020){Lutz}, {Sturm}, {Janssen}, {Veilleux}, {Aalto},
  {Cicone}, {Contursi}, {Davies}, {Feruglio}, {Fischer}, {Fluetsch},
  {Garcia-Burillo}, {Genzel}, {Gonz{\'a}lez-Alfonso}, {Graci{\'a}-Carpio},
  {Herrera-Camus}, {Maiolino}, {Schruba}, {Shimizu}, {Sternberg}, {Tacconi}, \&
  {Wei{\ss}}}]{Lutz2020}
{Lutz}, D., {Sturm}, E., {Janssen}, A., {et~al.} 2020, \aap, 633, A134

\bibitem[{{Neff} {et~al.}(2003){Neff}, {Ulvestad}, \& {Campion}}]{Neff2003}
{Neff}, S.~G., {Ulvestad}, J.~S., \& {Campion}, S.~D. 2003, \apj, 599, 1043

\bibitem[{{Ohyama} {et~al.}(2015){Ohyama}, {Terashima}, \&
  {Sakamoto}}]{Ohyama2015}
{Ohyama}, Y., {Terashima}, Y., \& {Sakamoto}, K. 2015, \apj, 805, 162

\bibitem[{{Pereira-Santaella} {et~al.}(2010){Pereira-Santaella},
  {Alonso-Herrero}, {Rieke}, {Colina}, {D{\'{\i}}az-Santos}, {Smith},
  {P{\'e}rez-Gonz{\'a}lez}, \& {Engelbracht}}]{Pereira2010}
{Pereira-Santaella}, M., {Alonso-Herrero}, A., {Rieke}, G.~H., {et~al.} 2010,
  \apjs, 188, 447

\bibitem[{{Pereira-Santaella} {et~al.}(2011){Pereira-Santaella},
  {Alonso-Herrero}, {Santos-Lleo}, {Colina}, {Jim{\'e}nez-Bail{\'o}n},
  {Longinotti}, {Rieke}, {Ward}, \& {Esquej}}]{Pereira2011}
{Pereira-Santaella}, M., {Alonso-Herrero}, A., {Santos-Lleo}, M., {et~al.}
  2011, \aap, 535, A93

\bibitem[{{Pereira-Santaella} {et~al.}(2016){Pereira-Santaella}, {Colina},
  {Garc{\'{\i}}a-Burillo}, {Alonso-Herrero}, {Arribas}, {Cazzoli}, {Emonts},
  {Piqueras L{\'o}pez}, {Planesas}, {Storchi Bergmann}, {Usero}, \&
  {Villar-Mart{\'{\i}}n}}]{Pereira2016b}
{Pereira-Santaella}, M., {Colina}, L., {Garc{\'{\i}}a-Burillo}, S., {et~al.}
  2016, \aap, 594, A81

\bibitem[{{Pereira-Santaella} {et~al.}(2018){Pereira-Santaella}, {Colina},
  {Garc{\'{\i}}a-Burillo}, {Combes}, {Emonts}, {Aalto}, {Alonso-Herrero},
  {Arribas}, {Henkel}, {Labiano}, {Muller}, {Piqueras L{\'o}pez}, {Rigopoulou},
  \& {van der Werf}}]{Pereira2018}
{Pereira-Santaella}, M., {Colina}, L., {Garc{\'{\i}}a-Burillo}, S., {et~al.}
  2018, \aap, 616, A171

\bibitem[{{Pereira-Santaella} {et~al.}(2021){Pereira-Santaella}, {Colina},
  {Garc{\'\i}a-Burillo}, {Lamperti}, {Gonz{\'a}lez-Alfonso}, {Perna},
  {Arribas}, {Alonso-Herrero}, {Aalto}, {Combes}, {Labiano},
  {Piqueras-L{\'o}pez}, {Rigopoulou}, \& {van der Werf}}]{Pereira2021}
{Pereira-Santaella}, M., {Colina}, L., {Garc{\'\i}a-Burillo}, S., {et~al.}
  2021, \aap, 651, A42

\bibitem[{{Pereira-Santaella} {et~al.}(2024{\natexlab{a}}){Pereira-Santaella},
  {Garc{\'\i}a-Bernete}, {Gonz{\'a}lez-Alfonso}, {Alonso-Herrero}, {Colina},
  {Garc{\'\i}a-Burillo}, {Rigopoulou}, {Arribas}, \& {Perna}}]{Pereira2024Mg4}
{Pereira-Santaella}, M., {Garc{\'\i}a-Bernete}, I., {Gonz{\'a}lez-Alfonso}, E.,
  {et~al.} 2024{\natexlab{a}}, \aap, 685, L13

\bibitem[{{Pereira-Santaella} {et~al.}(2024{\natexlab{b}}){Pereira-Santaella},
  {Gonz{\'a}lez-Alfonso}, {Garc{\'\i}a-Bernete}, {Donnan}, {Santa-Maria},
  {Goicoechea}, {Lamperti}, {Perna}, \& {Rigopoulou}}]{Pereira2024H3p}
{Pereira-Santaella}, M., {Gonz{\'a}lez-Alfonso}, E., {Garc{\'\i}a-Bernete}, I.,
  {et~al.} 2024{\natexlab{b}}, \aap, 689, L12

\bibitem[{{Pereira-Santaella} {et~al.}(2024{\natexlab{c}}){Pereira-Santaella},
  {Gonz{\'a}lez-Alfonso}, {Garc{\'\i}a-Bernete}, {Garc{\'\i}a-Burillo}, \&
  {Rigopoulou}}]{Pereira2024CO}
{Pereira-Santaella}, M., {Gonz{\'a}lez-Alfonso}, E., {Garc{\'\i}a-Bernete}, I.,
  {Garc{\'\i}a-Burillo}, S., \& {Rigopoulou}, D. 2024{\natexlab{c}}, \aap, 681,
  A117

\bibitem[{{Perna} {et~al.}(2024){Perna}, {Arribas}, {Lamperti},
  {Pereira-Santaella}, {Ulivi}, {B{\"o}ker}, {Maiolino}, {Bunker}, {Charlot},
  {Cresci}, {Rodr{\'\i}guez Del Pino}, {D'Eugenio}, {{\"U}bler}, {Fahrion}, \&
  {Ceci}}]{Perna2024}
{Perna}, M., {Arribas}, S., {Lamperti}, I., {et~al.} 2024, \aap, 690, A171

\bibitem[{{Ricci} {et~al.}(2017){Ricci}, {Trakhtenbrot}, {Koss}, {Ueda}, {Del
  Vecchio}, {Treister}, {Schawinski}, {Paltani}, {Oh}, {Lamperti}, {Berney},
  {Gandhi}, {Ichikawa}, {Bauer}, {Ho}, {Asmus}, {Beckmann}, {Soldi},
  {Balokovi{\'c}}, {Gehrels}, \& {Markwardt}}]{Ricci2017_BAT}
{Ricci}, C., {Trakhtenbrot}, B., {Koss}, M.~J., {et~al.} 2017, \apjs, 233, 17

\bibitem[{{Rich} {et~al.}(2023){Rich}, {Aalto}, {Evans}, {Charmandaris},
  {Privon}, {Lai}, {Inami}, {Linden}, {Armus}, {Diaz-Santos}, {Appleton},
  {Barcos-Mu{\~n}oz}, {B{\"o}ker}, {Larson}, {Law}, {Malkan}, {Medling},
  {Song}, {U}, {van der Werf}, {Bohn}, {Brown}, {Finnerty}, {Hayward},
  {Howell}, {Iwasawa}, {Kemper}, {Marshall}, {Mazzarella}, {McKinney},
  {Muller-Sanchez}, {Murphy}, {Sanders}, {Soifer}, {Stierwalt}, \&
  {Surace}}]{Rich2023}
{Rich}, J., {Aalto}, S., {Evans}, A.~S., {et~al.} 2023, \apjl, 944, L50

\bibitem[{{Rieke} {et~al.}(2025){Rieke}, {Buiten}, {Goldberg}, {Morrison}, {van
  der Werf}, {Alonso-Herrero}, {Alberts}, {Bonaventura}, {Ji}, {Lyu},
  {Rinaldi}, {Stone}, {Sun}, \& {Zhu}}]{Rieke2025}
{Rieke}, G.~H., {Buiten}, V.~A., {Goldberg}, C.~E., {et~al.} 2025, \apj, 988,
  17

\bibitem[{{Rigopoulou} {et~al.}(2024){Rigopoulou}, {Donnan},
  {Garc{\'\i}a-Bernete}, {Pereira-Santaella}, {Alonso-Herrero}, {Davies},
  {Hunt}, {Roche}, \& {Shimizu}}]{Rigopoulou2024}
{Rigopoulou}, D., {Donnan}, F.~R., {Garc{\'\i}a-Bernete}, I., {et~al.} 2024,
  \mnras, 532, 1598

\bibitem[{{Rodr{\'\i}guez Montero} {et~al.}(2019){Rodr{\'\i}guez Montero},
  {Dav{\'e}}, {Wild}, {Angl{\'e}s-Alc{\'a}zar}, \&
  {Narayanan}}]{RodriguezMontero2019}
{Rodr{\'\i}guez Montero}, F., {Dav{\'e}}, R., {Wild}, V.,
  {Angl{\'e}s-Alc{\'a}zar}, D., \& {Narayanan}, D. 2019, \mnras, 490, 2139

\bibitem[{{Sakamoto} {et~al.}(2014){Sakamoto}, {Aalto}, {Combes}, {Evans}, \&
  {Peck}}]{Sakamoto2014}
{Sakamoto}, K., {Aalto}, S., {Combes}, F., {Evans}, A., \& {Peck}, A. 2014,
  \apj, 797, 90

\bibitem[{{Sanders} \& {Mirabel}(1996)}]{Sanders96}
{Sanders}, D.~B. \& {Mirabel}, I.~F. 1996, \araa, 34, 749

\bibitem[{{Snyder} {et~al.}(2013){Snyder}, {Hayward}, {Sajina}, {Jonsson},
  {Cox}, {Hernquist}, {Hopkins}, \& {Yan}}]{Snyder2013}
{Snyder}, G.~F., {Hayward}, C.~C., {Sajina}, A., {et~al.} 2013, \apj, 768, 168

\bibitem[{{Solomon} \& {Vanden Bout}(2005)}]{Solomon2005}
{Solomon}, P.~M. \& {Vanden Bout}, P.~A. 2005, \araa, 43, 677

\bibitem[{Speranza} {et~al.}(2026){Pereira-Santaella}, {Garc{\'\i}a-Bernete}, {Gonz{\'a}lez-Alfonso}]{Speranza26}
{Speranza}, G., {Pereira-Santaella}, M., {Garc{\'\i}a-Bernete}, I., {et~al.} 2026, \aap,  accepted, arXiv:2610.24069

\bibitem[{{Steinacker} {et~al.}(2013){Steinacker}, {Baes}, \&
  {Gordon}}]{Steinacker2013}
{Steinacker}, J., {Baes}, M., \& {Gordon}, K.~D. 2013, \araa, 51, 63

\bibitem[{{Sturm} {et~al.}(2011){Sturm}, {Gonz{\'a}lez-Alfonso}, {Veilleux},
  {Fischer}, {Graci{\'a}-Carpio}, {Hailey-Dunsheath}, {Contursi}, {Poglitsch},
  {Sternberg}, {Davies}, {Genzel}, {Lutz}, {Tacconi}, {Verma}, {Maiolino}, \&
  {de Jong}}]{Sturm2011}
{Sturm}, E., {Gonz{\'a}lez-Alfonso}, E., {Veilleux}, S., {et~al.} 2011, \apjl,
  733, L16

\bibitem[{{van der Tak} {et~al.}(2007){van der Tak}, {Black}, {Sch{\"o}ier},
  {Jansen}, \& {van Dishoeck}}]{vanderTak2007}
{van der Tak}, F.~F.~S., {Black}, J.~H., {Sch{\"o}ier}, F.~L., {Jansen}, D.~J.,
  \& {van Dishoeck}, E.~F. 2007, \aap, 468, 627

\bibitem[{{Veilleux} {et~al.}(2013){Veilleux}, {Mel{\'e}ndez}, {Sturm},
  {Gracia-Carpio}, {Fischer}, {Gonz{\'a}lez-Alfonso}, {Contursi}, {Lutz},
  {Poglitsch}, {Davies}, {Genzel}, {Tacconi}, {de Jong}, {Sternberg}, {Netzer},
  {Hailey-Dunsheath}, {Verma}, {Rupke}, {Maiolino}, {Teng}, \&
  {Polisensky}}]{Veilleux2013}
{Veilleux}, S., {Mel{\'e}ndez}, M., {Sturm}, E., {et~al.} 2013, \apj, 776, 27

\bibitem[{{Yang} {et~al.}(2010){Yang}, {Stancil}, {Balakrishnan}, \&
  {Forrey}}]{Yang2010}
{Yang}, B., {Stancil}, P.~C., {Balakrishnan}, N., \& {Forrey}, R.~C. 2010,
  \apj, 718, 1062

\bibitem[{{Zepf} {et~al.}(1999){Zepf}, {Ashman}, {English}, {Freeman}, \&
  {Sharples}}]{Zepf1999}
{Zepf}, S.~E., {Ashman}, K.~M., {English}, J., {Freeman}, K.~C., \& {Sharples},
  R.~M. 1999, \aj, 118, 752

\end{thebibliography}
\end{document}